\documentclass[lettersize,journal]{IEEEtran}
\IEEEoverridecommandlockouts
\usepackage{cite}
\usepackage{amsmath,amssymb,amsfonts}
\usepackage{graphicx}
\usepackage{textcomp}
\usepackage{xcolor}
\usepackage{subcaption}
\usepackage{bm}
\usepackage{multirow}
\usepackage{booktabs}
\usepackage{subcaption}
\usepackage{amsthm}\usepackage{balance}\usepackage{url}
\usepackage{algorithm,algpseudocode}
\usepackage{amssymb}
\usepackage{pifont}

\newtheorem*{proof*}{Proof}

\algrenewcommand\algorithmicrequire{\textbf{Initialization:}}
\algrenewcommand\algorithmicensure{\textbf{Output:}}
\algdef{SE}[DOWHILE]{Do}{doWhile}{\algorithmicdo}[1]{\algorithmicwhile\ #1}%

\makeatletter

\newcommand{\Rmnum}[1]{\expandafter\@slowromancap\romannumeral #1@}
\makeatother

\date{}

\def\BibTeX{{\rm B\kern-.05em{\sc i\kern-.025em b}\kern-.08em
    T\kern-.1667em\lower.7ex\hbox{E}\kern-.125emX}}

\begin{document}
\title{Digital Semantic Communications: An Alternating Multi-Phase Training Strategy with Mask Attack
}
\author{Mingze Gong,~\IEEEmembership{Graduate Student Member,~IEEE}, Shuoyao Wang,~\IEEEmembership{Senior Member,~IEEE}, Suzhi Bi,~\IEEEmembership{Senior Member, IEEE}, 
Yuan Wu,~\IEEEmembership{Senior Member,~IEEE} and Liping Qian,~\IEEEmembership{Senior Member,~IEEE}
}

\maketitle
\begin{abstract}\label{abstract}
    Semantic communication (SemComm) has emerged as new paradigm shifts.
    Most existing SemComm systems transmit continuously distributed signals in analog fashion.
    However, the analog paradigm is not compatible with current digital communication frameworks. 
    In this paper, we propose an alternating multi-phase training strategy (AMP) to enable the joint training of the networks in the encoder and decoder through non-differentiable digital processes. 
    AMP contains three training phases, aiming at feature extraction (FE), robustness enhancement (RE), and training-testing alignment (TTA), respectively.
    In particular, in the FE stage, we learn the representation ability of semantic information by end-to-end training the encoder and decoder in an analog manner.
    When we take digital communication into consideration, the domain shift between digital and analog demands the fine-tuning for encoder and decoder. 
    To cope with joint training process within the non-differentiable digital processes, we propose the alternation between updating the decoder individually and jointly training the codec in RE phase.
    To boost robustness further, we investigate a mask-attack (MATK) in RE to simulate an evident and severe bit-flipping effect in a differentiable manner. 
    To address the training-testing inconsistency introduced by MATK, we employ an additional TTA phase, fine-tuning the decoder without MATK. 
    Combining with AMP and an information restoration network, we propose a digital SemComm system for image transmission, named AMP-SC\footnote{The code is available in https://github.com/gmzSZU/AMP-SC}. 
    Comparing with the representative benchmark, AMP-SC achieves $0.82 \sim 1.65$dB higher average reconstruction performance among various representative datasets at different scales and a wide range of signal-to-noise ratio. 

\end{abstract}
\section{Introduction}\label{introduction}
    \IEEEPARstart{s}{ixth} generation (6G) has been conceptualized as an intelligent information system both being driving and driven by the modern artificial intelligence (AI) technology \cite{letaief2019roadmap}. 
    Bridging the AI applications and physical world, SemComm achieves reliable transmission with impressive semantic level information fidelity \cite{gunduz2022beyond} by extracting the latent semantic features from the source. 
    In fact, SemComm has been recognized as a promising technique to improve communication efficiency and breakthrough beyond Shannon paradigm. 

    By leveraging the advances in deep learning, SemComm systems often use deep neural networks (DNNs) for joint source-channel coding (JSCC) to extract and encode the semantic information and can transmit various types of sources, such as texts\cite{xie2021deep}, images\cite{bourtsoulatze2019deep, zhang2023predictive}, speeches\cite{weng2021semantic}, and videos\cite{wang2022wireless}.
    For instance, \cite{xie2021deep} proposed a text transmission SemComm system called DeepSC, and \cite{bourtsoulatze2019deep} developed a SemComm system named DeepJSCC for image transmission. 
    Alternatively, in some cases, not all receivers are expected to display source data. 
    Instead, some receivers might be on duty of edge tasks, and they prefer directly inferring based on received semantic features rather than the reconstructed source data. 
    Motivated by this, numerous task-oriented SemComm systems \cite{shao2021learning, cai2023multi, zhang2022deep} have been proposed and demonstrated remarkable communication efficiency compared with separate source-channel coding (SSCC) approaches. 
    Besides of above-mentioned works, some works have investigated the combination of conventional SemComm and widely-used communication technologies, including non-orthogonal multiple access \cite{8952876, 10158994}, multiple-input-multiple-output \cite{10278812, 10026795}, and broadcast communication \cite{10335608, 10437222}.


    Nonetheless, most of the existing SemComm systems suffer compatibility issues with modern wireless communication framework. 
    The DNN-based deep channel encoders typically output continuous signals, which are then directly transmitted to physical channel.
    This analog transmission approach, however, is challenging to implement in practice due to the non-ideal characteristics of hardware components, such as power amplifiers.
    Although the implementation of analog paradigm based SemComm might appear straightforward, bridging conventional SemComm and practical applications necessitates a comprehensive evolution of hardware, protocols, and communication operators, which can be extremely costly. 
    Therefore, it is also crucial to explore approaches to convert continuous signals into finite ones to improve compatibility of SemComm system to conventional framework, enabling a smooth transition between two paradigms.

    \begin{figure*}[!t]
        \centering
        \includegraphics[scale=0.4]{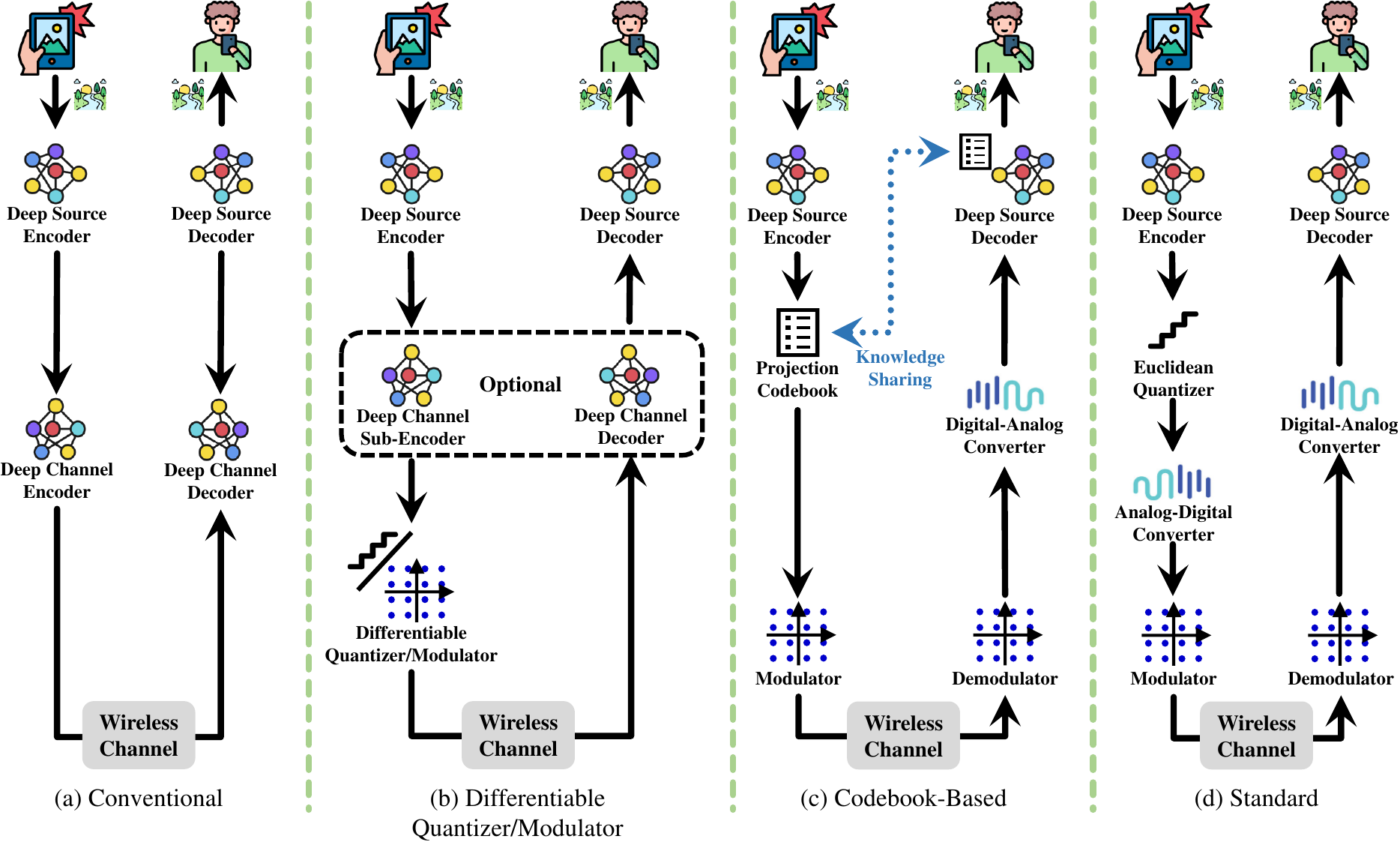}
        \caption{\small{Illustration of conventional SemComm system and various types of digital SemComm systems.
        Conventional denotes the conventional SemComm systems\cite{xie2021deep, bourtsoulatze2019deep, zhang2023predictive, weng2021semantic, wang2022wireless}, where the continuous semantic features are directly transmitted to the wireless channel. 
        Furthermore, Differentiable Quantizer/Modulator denotes the systems employed differentiable quantizers\cite{9953076,guo2023device,10505860,shao2022task, shao2023task} or learnable modulators \cite{tung2022deepjscc, bo2022learning} on the transmitter side. 
        Additionally, Codebook-Based \cite{xie2023robust, fu2023vector, bao2024sdac, huh2024universal, zhou2024moc} represents those leveraging learnable codebook for vector quantization and standard modulator for constellation mapping. 
        Standard represents the systems, including \cite{huang2024d, gao2023adaptive, he2022robust} and \emph{ours}, have simultaneously studied quantization, modulation, and demodulation, which are important processes in standard wireless communication framework. 
        In this paper, we aim at proposing a training strategy to enable a smooth transition from (a) to (d) for any SemComm system, empowering the digital SemComm with the development of conventional SemComm. 
        }}
        \label{fig:related_works}
    \end{figure*}

    To be compatible with the fact that the current wireless hardware/protocol can only admit certain sets of channel inputs, many research efforts have been devoted to the quantization problem in SemComm system.
    The non-differentiable quantization renders directly applying the conventional SemComm methods in digital fashion.
    In particular, some researches \cite{9953076,guo2023device,10505860} have proposed differentiable quantizers through mathematical approximation. 
    On the other hand, some works \cite{shao2022task, shao2023task} investigated the use of additive analog noise, which is differentiable, to replace quantization during training, yielding remarkable performance during digital testing. 

    Apart from discrete and finite set channel input, the process of current digital communication requires further mapping the transmitting signals into  constellation symbols, i.e., digital modulation.
    The constellation mapping ensures efficient modulation-demodulation, which are crucial for minimizing transmission errors in real-world noisy wireless channels.
    To cope with the non-differentiable issues caused by both the discrete-value-to-constellation mapping and the likelihood maximization during demodulation, \cite{tung2022deepjscc, bo2022learning} proposed learnable modulators to bypass conventional digital processes for end-to-end training. 
    That is, they empower the projection from continuous features to bit sequences by DNNs in the transmitter and enable the receiver to directly decode the received constellation symbols without likelihood maximization. 

    Although the learnable modulators have addressed the incompatibility in the transmitter, the receiver without demodulation is still in an analog fashion. 
    Accordingly, based on codebooks for quantization, some works \cite{xie2023robust, fu2023vector, bao2024sdac, huh2024universal} have explored approaches to train the networks through modulation and demodulation in an end-to-end manner. 
    In particular, \cite{xie2023robust} leveraged categorical reparameterization with Gumbel-Softmax to approximate the demodulation. 
    Alternatively, \cite{fu2023vector} adopt the gradient estimator to numerically approximate the gradient of likelihood maximization. 
    Furthermore, \cite{bao2024sdac} introduced the binary symmetric channel model to enable the gradient propagation. 
    However, the learnable modulators fit to specific modulation orders during training, leading to inflexibility in changing different modulations (e.g., QPSK, 16-QAM) according to channel conditions. 
    Thus, \cite{huh2024universal} proposed a modulation-agnostic SemComm framework that incorporates all the digital modulation orders in training. 
    However, the above codebook-based quantization methods \cite{xie2023robust, fu2023vector, bao2024sdac, huh2024universal} suffer the mismatch issue occurring in the local relationship between code indices and code vectors.    
    To address this issue, \cite{zhou2024moc} proposed a heuristic codebook reordering algorithm, where the Gray code mapping is integrated. 
    Despite, this algorithm aligns the codebook and modulation by establishing proximity relationships under the context of Gray code mapping, rather than together with the codebook. 
    Consequently, the learned semantic information space becomes distorted, resulting in the average performance degradation. 
    In brief, the mismatch issue between code indices and code vectors continues to undermine the effectiveness of codebook-based quantization methods.

    Apart from these codebook-based methods, the encoders in \cite{choi2019neural, park2023joint, song2020infomax} generate representations directly in binary latent space.
    The binary-latent approaches facilitate compatibility with existing modulation-demodulation schemes.
    However, the binary latent space is much larger than the discrete value space, since multiple bits are grouped to represent one discrete value. 
    Furthermore, when amounts of bits are transmitted for high-quality images, the binary DNNs could suffer the convergence problem in training and performance degradation in testing due to the explosively increased number of parameters.
    Accordingly, the proposed systems in \cite{choi2019neural, park2023joint, song2020infomax} only work for a short range of compression ratios. 
    In addition, it is worth noting that DNNs are widely researched based on the assumption of latent space in the field of real numbers.
    Hence, encoding the entries of feature vectors at the bit level is often inefficient and diminishes the benefits associated with SemComm\cite{huh2024universal}. 

    \begin{table}[!t]
        \centering
        \caption{\small{Considered Digital Processes of Related Works and Ours}}
        \resizebox{\linewidth}{!}{
        \begin{tabular}{cccc}
        \hline
        SemComm System   & Euclidean Quantization & Modulation  & Demodulation \\
        \hline
        \cite{9953076, guo2023device, 10505860, shao2022task, shao2023task} & \ding{51}     & \ding{55}     & \ding{55} \\
        \cite{tung2022deepjscc, bo2022learning} & \ding{55}     & \ding{51}     & \ding{55} \\
        \cite{xie2023robust, fu2023vector, bao2024sdac, huh2024universal, zhou2024moc} & \ding{55}     & \ding{51}     & \ding{51} \\
        \cite{huang2024d, gao2023adaptive, he2022robust}, \textbf{Ours} & \ding{51}     & \ding{51}     & \ding{51} \\
        \hline
        \end{tabular}%
        }
        \label{tab:digital_gap}%
    \end{table}%

    The aforementioned works have made impressive efforts to transfer continuous data streams to discrete data streams within the transmitted signal, achieving remarkable performance. 
    However, it is worth noting that a modern standard digital communication framework encompasses multiple digital processes, including source coding, channel coding, Euclidean quantization (e.g. rounding), modulation-demodulation, and analog-digital conversion. 
    To be more compatible with the standard framework, the JSCC-based SemComm system needs to consider all these digital processes, from Euclidean quantization to digital-analog conversion. 
    Unfortunately, as shown in Fig.~\ref{fig:related_works} and Table.~\ref{tab:digital_gap}, \emph{these works deviate from standard digital communication frameworks in one or more signal processing steps during conversion between continuous and finite signals.}
    This variance still hinders a smooth transition to semantic communication in a cost-effective manner.

    Recently, to facilitate the deployment of SemComm within standard digital communication frameworks, some works \cite{huang2024d, gao2023adaptive, he2022robust} have conducted further investigations on optimizing DNNs in the encoder and decoder through non-differentiable digital operations and channel.
    To enable the end-to-end training in digital SemComm system, \cite{huang2024d} proposed a digital SemComm system, where the source codecs and physical channel are modeled as a probabilistic model, for image transmission.
    Despite remarkable performance, the probabilistic approach imposes constraints on the upper bound of reconstruction performance due to the uncertainty \cite{valdenegro2021find}.

    To efficiently utilize powerful models for remarkable task performance in digital SemComm, \cite{gao2023adaptive, he2022robust} have explored decoder-only training strategy. 
    In particular, to avoid non-differentiability issue in modulation, \cite{gao2023adaptive} focused on optimizing networks in the receiver, rather than those in the transmitter.
    Similarly, \cite{he2022robust} proposed a two-phase training strategy.
    In the first phase, both the encoder and decoder are offline-trained in an end-to-end manner.
    In the second phase, \cite{he2022robust} solely fine-tuned the decoder, where the inputs are the received demodulated symbols generated by the frozen encoder with the conventional modulation-demodulation process. 
    Consequently, the encoder displays limited contribution to the network optimization, resulting in sub-optimal inference accuracy in the receiver. 
    Overall, \emph{the non-differentiability problem in digital SemComm system during training remains unsolved.}

    In this paper, towards the above challenge, we propose the alternating multi-phase training strategy (AMP) to overcome non-differentiability in training. 
    \emph{It is worth noting that the alternating training strategy can be applied to various SemComm systems for compatible implementation in current digital communication framework without any modification on neural network architectures.
    }In addition, AMP enables the noise robustness enhancement in both the encoder and decoder, improving the task performance in noisy channels. 
    Furthermore, we explore a differentiable mask-attack (MATK) to indicate the unpredictable noise during transmission. 
    From the perspective of DNN, we propose an information restoration network, called IRSNet, to reduce the impact of bit-flipping (BFP) in the receiver. 
    Additionally, based on JSCC framework, we propose a digital semantic communication system, named AMP-SC, for image transmission. 
    Notably, standard modulation schemes such as QPSK and 16QAM can be directly implemented in AMP-SC, which is empowered by AMP.
    The major contributions of this paper are summarized as follows:
    
    \begin{enumerate}
        \item {\emph{Alternating Multi-Phase Training Strategy}}: Given the non-differentiable nature of bits transmission, we propose a training strategy named AMP to optimize the system, empowering the success of conventional SemComm into modern digital communication frameworks. 
        \item We design AMP with three training phases: feature extraction, robustness enhancement, and training-testing alignment, respectively. In the robustness enhancement phase, we introduce a differentiable MATK to approximate and indicate the unpredictable noise during transmission, thereby boosting the noise robustness.
        \item{\emph{IRSNet}}: Inspired by the great success of U-Net in signal/image restoration, we develop IRSNet to restore noise-corrupted semantic features to cleaner ones, mitigating the impact of unpredictable BFP for the decoder.
        \item {\emph{Superior Performance} }: Compared with the state-of-the-art approaches, AMP-SC achieves remarkable reconstruction performance. In terms of 4-bit quantization, AMP-SC displays $1.24 \sim 1.65$dB higher average performance than the digital system empowered by the ``soft'' quantization in \cite{shao2023task} and two-phase training strategy in \cite{he2022robust} among multiple representative datasets.
    \end{enumerate}

    The remainder of this paper is organized as follows: In the following Section~\ref{proposed_method}, we will introduce the system model of the proposed AMP-SC.
    In particular, we will discuss the proposed AMP and neural network architecture in details in Section~\ref{AMP}. 
    Then, simulation results are presented in Section~\ref{simulation_results}.
    Finally, we will summarize this paper in Section~\ref{conclusion}. 
\section{System Model}\label{proposed_method}
  In this section, we will introduce the system model of the proposed AMP-SC.
  As shown in Fig.~\ref{fig:system_model}, we consider a single-user digital semantic communication system for image transmission. 
  Without loss of generality, we denote that the inputs of the SemComm system are images $\bm{x} \in \mathbb{R}^{c \times h \times w}$ with the number of color channels $c$ as well as the height $h$ and the width $w$ of the image signal and the outputs are the reconstructed images $\bm{y} \in \mathbb{R}^{c \times h \times w}$. 
\subsection{Transmitter}
  As shown in Fig.~\ref{fig:system_model}, the transmitter contains an encoder, a rounding quantizer, an analog-digital converter, and a modulator. 
  The encoder extracts features through learned function $\text{E}_{\mathrm{S}} :: \mathbb{R}^{c \times h \times w} \to \mathbb{R}^{n}$, where $n$ is the number of transmitted continuous symbols.
  Briefly, the deep source encoding function is given by 
  \begin{equation}
    \tilde{\bm{x}} = \text{E}_{\mathrm{S}}(\bm{x}|\bm{\phi}), \label{semantic_encoding}
  \end{equation}
  where $\tilde{\bm{x}} \in \mathbb{R}^{n}$ is the continuous encoded semantic features and $\bm{\phi}$ is the set of learnable parameters of the encoder. 
  Thus, the following quantized continuous semantic features can be represented as $\lceil \tilde{\bm{x}} \rfloor$, where $\lceil \cdot \rfloor$ is a rounding operation. 
  Assuming the input continuous symbol is $s$, the conversion from continuous symbols to binary bits can be achieved by following 
  \begin{subequations}
    \label{AD}
		\begin{align}
			\text{AD}(s) &= d_{b-1}d_{b-2}\ldots d_{1}d_{0},\\
			d_i &= \lfloor\frac{s}{2^{i}}\rfloor \bmod 2, 
 		\end{align}
	\end{subequations}
  where $d_i$ denotes the $i$-th bit counting from the right and $b$ is the number of bits per symbol for modulation.
  In addition, $\lfloor \cdot \rfloor$ indicates operation of rounding down. 
  Following the employed general constellation mapping and modulation schemes, the binary signal $\bm{z} \in \mathbb{C}^{n \times 2^{b}}$ are transmitted to physical channel. 

\subsection{Modulation, Physical Channel, and Demodulation}
  Due to the impact of physical channel noise, the transmitted bit stream will inevitably experience bit-flipping.
  This means that a bit transmitted as 0 (or 1) may be received as 1 (or 0).
  The probability of each bit-flipping and the correlation between adjacent flipped bits are influenced not only by the intensity of channel noise but also by the specific modulation and demodulation methods\cite{he2022robust}.
  Modeling the joint probability distribution of bit-flipping for each modulation-demodulation method is very complex, and training a model for each method individually is computationally expensive and beyond the scope of this paper. 
  Alternatively, this paper aims to use a simple yet representative model to analyze changes within a bit stream. 
  This approach leverages the simplicity of network optimization and extends to practical communication applications. 
  In particular, the binary symmetric channel model is widely adopted as the channel model to analyze bit transmission \cite{park2023joint}. 
  Following \cite{guo2023device,bao2024sdac,he2022robust, choi2019neural, park2023joint, song2020infomax}, we consider the binary symmetric channel as the channel model due to its symmetric error characteristics, making system optimization and performance validation more efficiently. 

\begin{figure}[!t]
    \centering
    \includegraphics[scale=0.4]{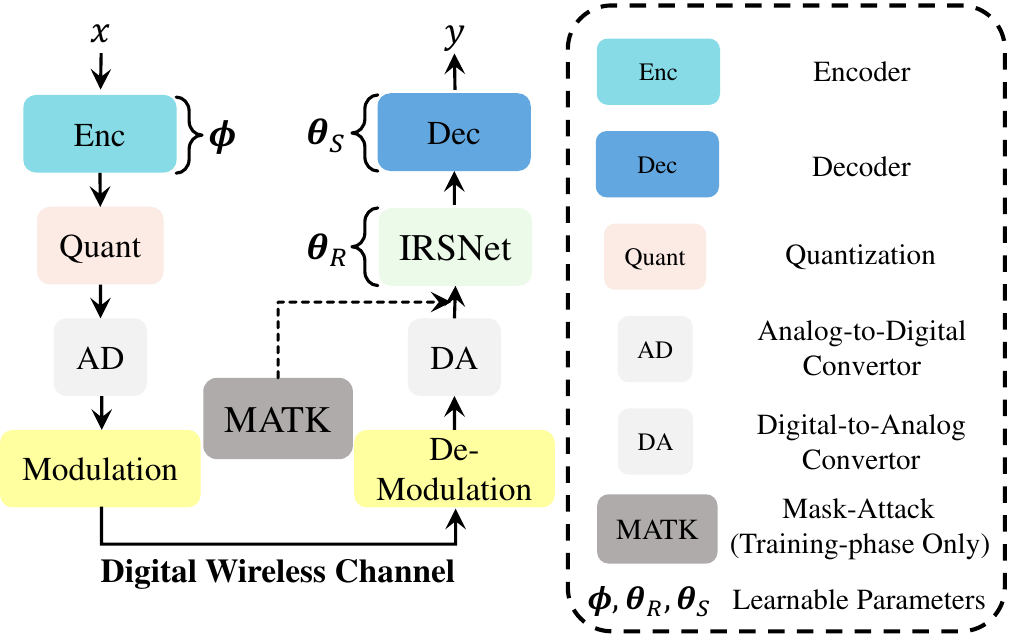}
    \caption{\small{System model of the proposed AMP-SC.}}
    \label{fig:system_model}
  \end{figure}

  As the bits are transmitted to the physical channel, some factors, such as channel noise, interference, and multipath effects, can lead to changes in the transmitted bits.
  Hence, the received bits could be inconsistent with the transmitted ones. 
  Despite the advanced maturity of modern error detection and correction techniques, they are still not flawless enough to detect and correct all errors.
  Therefore, from an end-to-end perspective of the converters in Fig.~\ref{fig:system_model}, some bits are flipped imperceptibly, known as BFP.
  Similar with \cite{song2020infomax}, we assume that BFP occurs independently in each bit and the probability model can be regarded as a binomial distribution with flipping probability $p$. 
  Therefore, the BFP can be given by 
  \begin{equation}
    \hat{\bm{z}} = \bm{z} + \text{sign}(\text{rand}(\bm{z})), \label{bit_flipping}
  \end{equation}
  where $\hat{\bm{z}} \in \mathbb{C}^{n\times2^{b}}$ denotes the received binary signal.
  Given an input $a$, the sign function $\text{sign}()$ follows 
  \begin{equation}
    \text{sign}(a) = 
    \begin{cases}
      1 & \text{if } a<p \\
      0 & \text{if } a>p
    \end{cases}, \label{sign}
  \end{equation}
  In addition, $\text{rand}(\bm{z}) \in \mathbb{C}^{n\times2^{b}}$ obeys the uniform distribution of $[0, 1]$ and is utilized to determine whether each bit in $\bm{z}$ is flipped or not.  

\subsection{Receiver}
  Upon receiving the corrupted bit-stream and converting it to continuous symbols, the receiver reconstructs the source information with neural networks. 
  However, BFP, as mentioned earlier, is unpredictable, making it inadequate to directly input the received continuous signal to the decoder as it may contain incorrect semantic information.  
  Therefore, we suggest restoring the received continuous signal with a DNN and then reconstruct the source data. 

  Similar with the transmitter, the receiver consists of the demodulator, digital-analog converter, optimizable information restoration network, and the decoder. 
  The binary semantic features are converted to the continuous ones $\check{\bm{z}} \in \mathbb{R}^{n}$ according to a digital-to-analog conversion function $\text{DA}(\cdot)$.
  Given the demodulated binary number set $\hat{\bm{d}}$, each grouped of demodulated digital representation in the received signal is first converted to continuous symbol by following
  \begin{equation}
    \text{DA}(\hat{\bm{d}}) = \sum_{i=0}^{b} (\hat{\bm{d}}_i \times 2^i). \label{DA}
  \end{equation}
  In particular, we leverage an information restoration network to reduce information loss, which is mainly caused by bit-flipping, for the decoder. 
  The restored semantic features $\tilde{\bm{z}} \in \mathbb{R}^{n}$ is given by 
  \begin{equation}
        \tilde{\bm{z}} = \text{D}_{\mathrm{R}}(\check{\bm{z}}|\bm{\theta}_{\mathrm{R}}),
  \end{equation}
  where $\text{D}_{\mathrm{R}} \in \mathbb{R}^{n} \to \mathbb{R}^{n}$ is the restoration function with optimizable parameters $\bm{\theta}_{\mathrm{R}}$. 
  Accordingly, the inference result is given by 
  \begin{equation}
    \bm{y} = \text{D}_{\mathrm{S}}(\tilde{\bm{z}}|\bm{\theta}_\mathrm{S}),  
  \end{equation}
  where $\text{D}_\mathrm{S} \in \mathbb{R}^{n} \to \mathbb{R}^{c \times h \times w}$ is the decoding function with learnable parameters $\bm{\theta}_\mathrm{S}$. 

  \begin{figure*}[!t]
    \centering
    \includegraphics[scale=0.45]{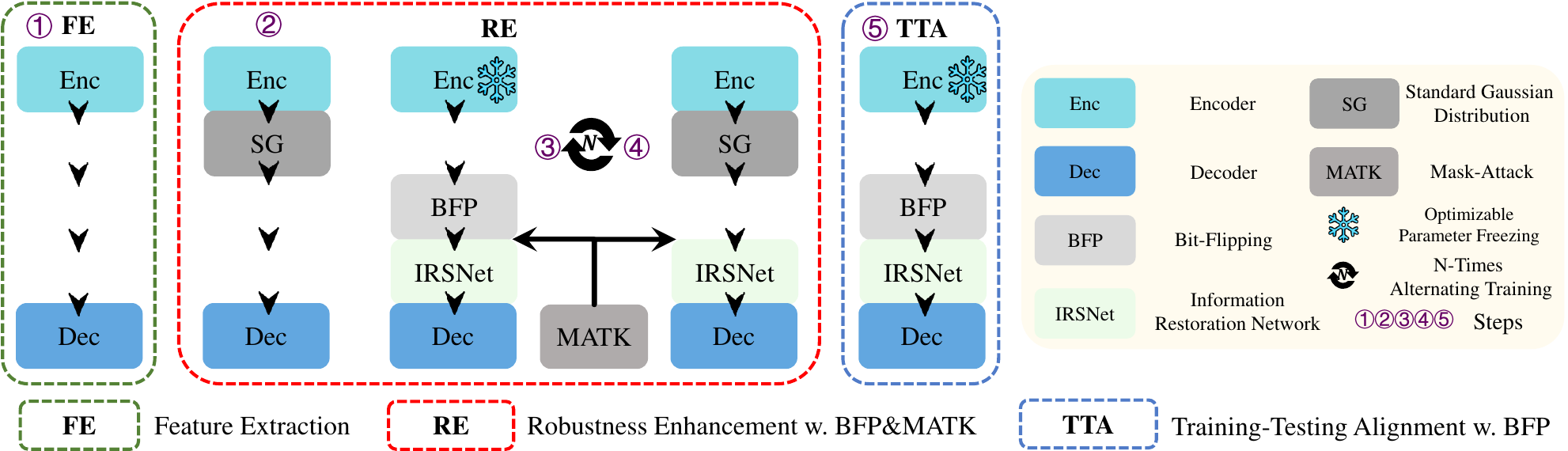}
    \caption{\small{The alternating multi-phase training processes for AMP-SC.}}
    \label{fig:pipeline}
  \end{figure*}  

\section{Alternating Multi-Phase Training Strategy and Network Architecture}\label{AMP}
  In this section, we will introduce the proposed training strategy. 
  As discussed in Section~\ref{introduction}, the decoding-only training strategy proposed by \cite{he2022robust} enables parameter tuning in the decoder to fit with BFP in an end-to-end manner. 
  However, the encoder remains in a noise-unaware fashion, since the gradients from the decoder are blocked by quantization and modulation. 
  The concession degrades the inference accuracy over the noisy wireless channel. 
  To tackle the non-differentiability in training and robustness enhancement for digital SemComm systems, we propose an alternating multi-phase training strategy, named AMP. 
  As shown in Fig.~\ref{fig:pipeline}, the proposed method mainly contains three phases, aiming at feature extraction (FE), robustness enhancement (RE), and training-testing alignment (TTA). 
  While each of FE and TTA contains 1 step, RE contains 3 steps, including 2 steps for alternating training. 
  AMP can be briefly described as the following:
  \begin{itemize}
    \item In the first phase FE, we aim at empowering the representation ability of the system in the analog domain. 
    \item Following FE, RE is deployed to intensify the robustness to information loss, which is caused by quantization and BFP, with ``soft'' quantization and alternating training. 
    In particular, the alternation between solely updating decoder and jointly optimizing codecs in RE, enables the parameter updating in both encoder and decoder, even the digital communication processes in-between are non-differentiable. 
    To boost robustness further, we investigate MATK in RE to simulate an evident and severe BFP effect in a differentiable manner. 
    \item To cope with the training-testing inconsistency introduced by MATK, we employ an additional TTA phase, fine-tuning the decoder without MATK. 
  \end{itemize}

  
  Algorithm~\ref{algorithm} briefly demonstrates the entire training procedures of AMP. 
  The training settings of AMP for AMP-SC are shown in Table.~\ref{tab:training_settings_ours}. 
  Particularly, in the following, $\bm{y}_i$ will be used to denote the reconstructed images in Step-$i$ for simplicity. 
  Moreover, $\mathcal{L}(\cdot)$ is used to represent the loss function, whose inputs are learnable parameters. 
  \begin{algorithm}[!t]
    \caption{Alternating Multi-Phase Training Strategy}
    \label{algorithm}
    \begin{algorithmic}[1]
        \State \textbf{Input}: a set of numbers of training epochs, batch size $M$, number of bits for analog-digital conversion $B$, BFP ratio $p$, and mask ratio $2p$
        \State \textbf{do} Step-1 and Step-2
        \State \textbf{while} alternating training round $j=1$ to $3$ \textbf{do}
        \State \hspace{0.5cm} \textbf{do} Step-3 and Step-4
        \State \textbf{end while}
        \State \textbf{do} Step-5
    \end{algorithmic}
\end{algorithm}

  \subsection{Feature Extraction}
  For SemComm, either in digital or analog fashion, one key sector is to extract the semantic-relevant information and compress the irrelevant information.
  As discussed in Section~\ref{introduction}, despite the differentiability of systems in \cite{choi2019neural,park2023joint,song2020infomax} empowered by binary DNNs, the binary latent space limits the range of compression ratios. 
  Rather than directly training in the digital domain, we argue that \emph{it is more efficient to gradually transfer well-optimized parameters from the analog domain to the digital domain. }

  Therefore, we eliminate the process of quantization and modulation, and thus directly feed the output of the encoder as the input of the decoder in FE. 
  In this case, we enable the end-to-end training of $\phi$ and $\theta_s$ in the analog domain to obtain the encoder and decoder that successfully extract the semantic-relevant information and compress the irrelevant information.
  Briefly, FE contains Step-1 only, whose training procedures are shown in Algorithm~\ref{algorithm_step1}. 
  Recalling the encoded continuous semantic features $\tilde{\bm{x}}$ in (\ref{semantic_encoding}), we optimize the model in the analog domain using an end-to-end training approach, following:
  \begin{equation}
    \bm{y}_1 = \text{D}_{\mathrm{S}}\left(\tilde{\bm{x}}|\bm{\theta}_\mathrm{S}\right), \label{y1}
  \end{equation}
  and the loss function is given by 
  \begin{equation}
    \mathcal{L}\left(\bm{\phi}, \bm{\theta}_\mathrm{S}\right) = \Vert \bm{y}_1-\bm{x} \Vert_1. \label{loss1}
  \end{equation}

  \begin{algorithm}[!t]
    \caption{Step-1 in Alternating Multi-Phase Training}
    \label{algorithm_step1}
    \begin{algorithmic}[1]
        \State \textbf{Input}: number of training epochs $K_1$, batch size $M$, and number of bits for analog-digital conversion $B$
        \State \textbf{while} epoch $k=1$ to $K_1$ \textbf{do}
        \State \hspace{0.5cm} Select a mini-batch of data $\{\bm{x}_{m}\}^{M}_{m=1}$
        \State \hspace{0.5cm} Compute feature vectors based on (\ref{semantic_encoding})
        \State \hspace{0.5cm} Compute reconstructed source image $\{\bm{y}_{1,m}\}^{M}_{m=1}$ based on (\ref{y1})
        \State \hspace{0.5cm} Compute the loss based on (\ref{loss1})
        \State \hspace{0.5cm} Update parameters $\bm{\phi}, \bm{\theta_{\mathrm{S}}}$ through backpropagation
        \State \textbf{end while}
    \end{algorithmic}
\end{algorithm}

\subsection{Robustness Enhancement}
  Notably, in FE, the system is well optimized in the analog domain to ensure the representation ability.
  When we take digital communication into consideration, the domain shift between digital and analog demands the fine-tuning for encoder and decoder.
  Accordingly, in this phase, we aim at enhancing robustness to noise distortion in the digital domain.

  In particular, to be compatible with current digital communication framework, a SemComm system should take digital processes, from quantization to demodulation, into further consideration.
  However, shifting to digital communications, the non-differentiability problem inevitably appears. 
  Particularly, the conversion between bit-stream and continuous signal-stream is not differentiable, as indicated by (\ref{AD}).
  This characteristic makes the end-to-end training approach no longer suitable to fine-tune the digital SemComm system. 
  Moreover, although the two-phase training strategy proposed by \cite{he2022robust} has enabled the optimization in the decoder, the encoder remains in a noise-unaware fashion, resulting in performance degradation over noisy channel. 
  Therefore, introducing digital processes, we alternate between solely updating decoder and jointly optimizing codecs in RE.   
  In the following, we will introduce the Step-2, alternating Step-3 and Step-4, alternating training with MATK in RE. 
  Additionally, the proposed IRSNet and alternating training of AMP will be discussed.

\subsubsection{Methodology of Step-2}
  In order to fine-tune the encoder and decoder through non-differentiable quantization, inspired by \cite{shao2023task}, we fine-tune the well-trained model in FE with additive randomness to achieve relaxed continuous quantization in the analog domain. 
  It also helps reduce information loss caused by quantization.
  Thus, in Step-2, the deep source decoding function can be denoted as 
  \begin{equation}
    \bm{y}_2 = \text{D}_{S}\left(\tilde{\bm{x}}+\alpha \times\bm{\mathcal{N}}|\bm{\theta}_S\right), \label{y2}
  \end{equation}
  where $\bm{\mathcal{N}}$ samples from a standard Gaussian distribution and $\alpha$ is the intensity coefficient of $\bm{\mathcal{N}}$. 
  Thus, the loss function is given by 
  \begin{equation}
    \mathcal{L}\left(\bm{\phi}, \bm{\theta}_S\right) = \Vert \bm{y}_2-\bm{x} \Vert_1.\label{loss2}
  \end{equation}
  The training procedures of Step-2 is shown in Algorithm~\ref{algorithm_step2}. 

  \begin{algorithm}[!t]
    \caption{Step-2 in Alternating Multi-Phase Training}
    \label{algorithm_step2}
    \begin{algorithmic}[1]
        \State \textbf{Input}: number of training epochs $K_2$, batch size $M$, and number of bits for analog-digital conversion $B$
        \State \textbf{while} epoch $k=1$ to $K_2$ \textbf{do}
        \State \hspace{0.5cm} Select a mini-batch of data $\{\bm{x}_m\}^{M}_{m=1}$
        \State \hspace{0.5cm} Compute feature vectors based on (\ref{semantic_encoding})
        \State \hspace{0.5cm} Randomly sample $\bm{\mathcal{N}}$ from a standard Gaussian distribution
        \State \hspace{0.5cm} Compute reconstructed source image $\{\bm{y}_{2, m}\}^{M}_{m=1}$ based on (\ref{y2})
        \State \hspace{0.5cm} Compute the loss based on (\ref{loss2})
        \State \hspace{0.5cm} Update parameters $\bm{\phi}, \bm{\theta_{\mathrm{S}}}$ through backpropagation
        \State \textbf{end while}
    \end{algorithmic}
\end{algorithm}

  \subsubsection{Alternating Step-3 and Step-4}
  In the following, as the impact of information loss caused by non-differentiable quantization is mitigated in Step-2, the optimization is concentrated on adapting the networks to digital channel environment in the following alternating steps. 
  To better illustrate the proposed alternating training, we suggest treating the transmitter and physical channel as a single entity.
  When the encoder networks are frozen to allow no parameter updates, the optimization of decoder networks becomes feasible, since the inputs to the decoder are demodulated symbols. 
  This approach bypasses the non-differentiable nature of the digital components in the system.
  Notably, the random BFP distorts the received signal, enabling the decoder networks to become aware of digital noise through backpropagation. 
  As the receiver is fine-tuned for down-stream task, the transmitter needs to adjust the encoding policy to further improve inference accuracy and noise robustness. 
  Thus, alignment in the latent space of transceivers is carried out by jointly training networks of the encoder and decoder with additive noise in the analog domain.  
  This joint training extends noise-awareness to the encoder, enhancing the overall system's robustness.

  Accordingly, we update the decoder's parameters with digital processes while freezing those of the encoder, which can be referred to Step-3. 
  Recalling the additive randomness, we then jointly optimize encoder and decoder networks in the analog domain to align the encoder and decoder networks, which is named as Step-4. 
  The alternating training enables the gradients to bypass the non-differentiable part of the digital system, allowing the learnable parameters to be updated.
  Furthermore, they contribute to the reduction of performance degradation in digital SemComm systems. 
  \emph{It is worth noting that such alternating training strategy can be transferred to various SemComm systems for implementation in current digital communication framework without any modification on network architectures. } 

  However, it remains challenging to preserve the transmitted semantic information from BFP.
  Since it is nearly unpredictable which bits will be flipped and how much the corresponding symbols will change, we argue that limited optimization can be achieved by the transmitter, while the burden of eliminating ``fake'' semantic information falls on the receiver.
  As for the receiver, \emph{one effective method to mitigate the impact of BFP is to restore the received signal before reconstructing the source information by a DNN. 
  }Motivated by this, we investigate differentiable MATK in the alternating training process to enhance robustness to the potential information loss caused by BFP and propose IRSNet to restore the transmitted continuous signal to reduce the negative impact of BFP, respectively.  
  
  Inspired by \cite{he2022masked}, MATK is designed to simulate an evident and severe bit-flipping effect in a differentiable manner. 
  With MATK in Step-3, the decoder infers based on the incomplete signal, where the semantic information is broken by MATK and explicitly polluted by BFP. 
  Thus, the decoder needs to learn the deep correlations among semantic information, resulting in the improvement of noise robustness. 
  Additionally, the encoder can also benefit from MATK.
  In Step-4, MATK provides simulated digital communication environment to empower the joint training of the encoder and decoder in the analog domain. 
  As a result, MATK enhances the error resilience of transmitted semantic information.

  Briefly, RE includes Step-2 and two alternating steps, which are named as Step-3 and Step-4, to enhance the system robustness to digital noise distortion. 
  It is worth noting that Step-3 and Step-4 will be alternated for multiple times, whereas Step-2 will only be carried out once.
  In Step-3, we train the decoder only, with random BFP in fixed flipping ratio $p$. 
  Notably, the learnable parameters in transmitter are frozen. 
  Recalling the received continuous semantic features $\check{\bm{z}}$ and semantic decoding function $\text{D}_S$, the forward propagation in receiver is given by 
  \begin{equation}
    \bm{y}_3 = \text{D}_{\mathrm{S}}\left(\check{\bm{z}} \cdot \bm{\mathcal{M}}|\bm{\theta}_{\mathrm{S}}\right), \label{y3_amp}
  \end{equation}
  where $\bm{\mathcal{M}}$ is the randomly generated binary mask and $\cdot$ denotes dot-product operation. 
  The loss function is given by 
  \begin{equation}
    \mathcal{L}\left(\bm{\theta}_\mathrm{S}\right) = \Vert \bm{y}_3-\bm{x} \Vert_1. \label{loss3}
  \end{equation}
  In Step-4, we jointly train the encoder and decoder with relaxed quantization and MATK.
  The semantic decoding function is given by 
  \begin{equation}
    \bm{y}_4 = \text{D}_{\mathrm{S}}\left(\left(\tilde{\bm{x}}+\alpha \times\bm{\mathcal{N}}\right) \cdot \bm{\mathcal{M}}|\bm{\theta}_\mathrm{S}\right).  \label{y4_amp}
  \end{equation}
  Although the process functions are similar with those in Step-3, the loss function is different and given by 
  \begin{equation}
    \mathcal{L}\left(\bm{\phi}, \bm{\theta}_{\mathrm{S}}\right) = \Vert \bm{y}_4-\bm{x} \Vert_1. \label{loss4}
  \end{equation}


  \begin{algorithm}[!t]
    \caption{Step-3 in Alternating Multi-Phase Training}
    \label{algorithm_step3}
    \begin{algorithmic}[1]
        \State \textbf{Input}: number of training epochs $K_3$, batch size $M$, number of bits for analog-digital conversion $B$, BFP ratio $p$, and mask ratio $2p$
        \State \textbf{while} epoch $k=1$ to $K_3$ \textbf{do}
        \State \hspace{0.5cm} Select a mini-batch of data $\{\bm{x}_{m}\}^{M}_{m=1}$
        \State \hspace{0.5cm} Compute feature vectors based on (\ref{semantic_encoding})
        \State \hspace{0.5cm} Compute quantized feature vectors
        \State \hspace{0.5cm} Compute analog-digital conversion based on (\ref{AD})
        \State \hspace{0.5cm} Compute binary signal $\{\bm{z}_{m}\}^{M}_{m=1}$ based on a given modulation order
        \State \hspace{0.5cm} Compute noise-corrupted signal $\{\hat{\bm{z}}_{m}\}^{M}_{m=1}$ based on (\ref{bit_flipping}), (\ref{sign}), and (\ref{DA})
        \State \hspace{0.5cm} Randomly sample $\bm{\mathcal{M}}$ from a uniform distribution between 0 and 1
        \State \hspace{0.5cm} Compute reconstructed source image $\{\bm{y}_{3, m}\}^{M}_{m=1}$ based on (\ref{y3_amp})
        \State \hspace{0.5cm} Compute the loss based on (\ref{loss3})
        \State \hspace{0.5cm} Update parameters $\bm{\theta_{\mathrm{R}}}, \bm{\theta_{\mathrm{S}}}$ through backpropagation
        \State \textbf{end while}
    \end{algorithmic}
\end{algorithm}

  Generally, the mask ratio should be larger than the bit-flipping ratio $p$. 
  As indicated by \cite{zhang2022mask}, the optimal mask ratio for a network could be high. 
  Furthermore, a good mask ratio should be selected to balance the distance between irrelevant semantics and that between relative semantics, which contribute to noise robustness and representation ability respectively. 
  Accordingly, without loss of generality, we suggest setting the mask ratio as $2p$. 
\subsubsection{IRSNet}
  Inspired by the great success of U-Net \cite{ronneberger2015u} in signal/image restoration, we develop IRSNet to restore noise-corrupted semantic features to cleaner ones, mitigating the impact of unpredictable BFP for the decoder.
  The multi-level structure with full connections of U-Net endows the network with multi-scale information extraction and restoration capabilities. 
  \emph{Furthermore, MATK makes the IRSNet blind to a portion of the received information, forcing it to restore the broken semantics. 
  }Although BFP introduces ``fake'' semantic information before masking, it provides no assistance in repairing the compromised semantics to IRSNet as it represents nothing in the source data.
  Through backpropagation, the IRSNet learns to achieve restoration with accurate semantic information in the corrupted signal.
  Consequently, the following decoder can reconstruct the source information with cleaner and more precise semantic information. 
  
  \begin{algorithm}[!t]
    \caption{Step-4 in Alternating Multi-Phase Training}
    \label{algorithm_step4}
    \begin{algorithmic}[1]
        \State \textbf{Input}: number of training epochs $K_4$, batch size $M$, number of bits for analog-digital conversion $B$, BFP ratio $p$, and mask ratio $2p$
        \State \textbf{while} epoch $k=1$ to $K_4$ \textbf{do}
        \State \hspace{0.5cm} Select a mini-batch of data $\{\bm{x}_{m}\}^{M}_{m=1}$
        \State \hspace{0.5cm} Compute feature vectors based on (\ref{semantic_encoding})
        \State \hspace{0.5cm} Randomly sample $\bm{\mathcal{N}}$ from a standard Gaussian distribution
        \State \hspace{0.5cm} Randomly sample $\bm{\mathcal{M}}$ from a uniform distribution between 0 and 1
        \State \hspace{0.5cm} Compute reconstructed source image $\{\bm{y}_4\}^{B}_{b=1}$ based on (\ref{y4_amp})
        \State \hspace{0.5cm} Compute the loss based on (\ref{loss4})
        \State \hspace{0.5cm} Update parameters $\bm{\phi}, \bm{\theta_{\mathrm{R}}}, \bm{\theta_{\mathrm{S}}}$ through backpropagation
        \State \textbf{end while}
    \end{algorithmic}
\end{algorithm}

  \subsubsection{Alternating Training of AMP-SC}

  With the joining of IRSNet, the process and loss functions of alternating training need to be rewritten for AMP-SC. 
  Recalling the functions (\ref{y3_amp}) and (\ref{loss3}) in Step-3, now we train the proposed IRSNet and decoder with random BFP in fixed flipping ratio and MATK in fixed mask ratio. 
  With the restoration function $\text{D}_\mathrm{R}$, the forward propagation in receiver is given by 
  \begin{equation}
    \bm{y}_3 = \text{D}_{\mathrm{S}}\left(\text{D}_{\mathrm{R}}\left(\check{\bm{z}} \cdot \bm{\mathcal{M}}|\bm{\theta}_{\mathrm{R}}\right)|\bm{\theta}_{\mathrm{S}}\right). \label{y3_ampsc}
  \end{equation}
  In order to optimize the proposed IRSNet for information restoration, we leverage the L1-loss to train IRSNet. 
  Accordingly, the loss function is given by 
  \begin{equation}
    \mathcal{L}\left(\bm{\theta}_\mathrm{R}, \bm{\theta}_\mathrm{S}\right) = \Vert \bm{y}_3-\bm{x} \Vert_1 + \Vert \tilde{\bm{z}}-\lceil \tilde{\bm{x}} \rfloor \Vert_1. \label{loss3_ampsc}
  \end{equation}
  In the following, the deep source decoding function in Step-4 is given by 
  \begin{equation}
    \bm{y}_4 = \text{D}_{\mathrm{S}}\left(\text{D}_{\mathrm{R}}\left(\left(\tilde{\bm{x}}+\alpha \times\bm{\mathcal{N}}\right) \cdot \bm{\mathcal{M}}|\bm{\theta}_{\mathrm{R}}\right)|\bm{\theta}_{\mathrm{S}}\right).  \label{y4_ampsc}
  \end{equation}
  Similar with (\ref{loss4}), the loss function is given by 
  \begin{equation}
    \mathcal{L}\left(\bm{\phi}, \bm{\theta}_{\mathrm{R}}, \bm{\theta}_{\mathrm{S}}\right) = \Vert \bm{y}_4-\bm{x} \Vert_1. \label{loss4_ampsc}
  \end{equation}

  Since the IRSNet is introduced and randomly initialized in the first round of interactive training, the number of training epoch in Step-3 of the first round is set to $4T$, where $T$ is the number of training epochs for each regular step, to refine the IRSNet and decoder. 

  \begin{algorithm}[!t]
    \caption{Step-5 in Alternating Multi-Phase Training}
    \label{algorithm_step5}
    \begin{algorithmic}[1]
        \State \textbf{Input}: number of training epochs $K_5$, batch size $M$, number of bits for analog-digital conversion $B$, and BFP ratio $p$
        \State \textbf{while} epoch $k=1$ to $K_5$ \textbf{do}
        \State \hspace{0.5cm} Select a mini-batch of data $\{\bm{x}_{m}\}^{M}_{m=1}$
        \State \hspace{0.5cm} Compute feature vectors based on (\ref{semantic_encoding})
        \State \hspace{0.5cm} Compute quantized feature vectors
        \State \hspace{0.5cm} Compute analog-digital conversion based on (\ref{AD})
        \State \hspace{0.5cm} Compute binary signal $\{\bm{z}_{m}\}^{M}_{m=1}$ based on a given modulation order
        \State \hspace{0.5cm} Compute noise-corrupted signal $\{\hat{\bm{z}}_{m}\}^{M}_{m=1}$ based on (\ref{bit_flipping}), (\ref{sign}), and (\ref{DA})
        \State \hspace{0.5cm} Compute reconstructed source image $\{\bm{y}_{5, m}\}^{M}_{m=1}$ based on (\ref{y5})
        \State \hspace{0.5cm} Compute the loss based on (\ref{loss5})
        \State \hspace{0.5cm} Update parameters $\bm{\theta_{\mathrm{R}}}, \bm{\theta_{\mathrm{S}}}$ through backpropagation
        \State \textbf{end while}
    \end{algorithmic}
\end{algorithm}

  \subsection{Training-Testing Alignment}
  In the alternating part, MATK has enhanced the restoration ability of IRSNet and noise robustness of the system against BFP.
  However, it is noteworthy that \emph{MATK does not exist in practical communications. 
  }Moreover, further fine-tuning the IRSNet with MATK limits the improvement of restoration ability against BFP since less latent space of IRSNet is devoted to restoring ``fake'' information. 
  As a result, the reconstruction performance of the system degrades in implementation. 
  Therefore, in this phase, we aim at aligning the decoding process in the training and testing stages to improve performance in implementation.  
  Particularly, we remove MATK to bridge the gap between training and testing, making simulations more relevant to real transmission scenarios.
  In addition, it ensures that the system can focus on explicit semantic loss during subsequent fine-tuning processes. 


  Briefly, training-testing alignment phase (TTA) contains only one step, which is step 5. 
  Based on the above discussion, the deep source decoding function is given by 
  \begin{equation}
    \bm{y}_5 = \text{D}_{\mathrm{S}}\left(\text{D}_{\mathrm{R}}\left(\check{\bm{z}}|\bm{\theta}_{\mathrm{R}}\right)|\bm{\theta}_{\mathrm{S}}\right), \label{y5}
  \end{equation}
  Furthermore, the loss function is denoted as 
  \begin{equation}
    \mathcal{L}\left(\bm{\theta}_\mathrm{R}, \bm{\theta}_\mathrm{S}\right) = \Vert \bm{y}_5-\bm{x} \Vert_1 + \Vert \tilde{\bm{z}}-\lceil \tilde{\bm{x}} \rfloor \Vert_1. \label{loss5}
  \end{equation}

\subsection{Network Architecture} \label{network_architecture}
\begin{figure}[!t]
    \centering
    \includegraphics[scale=0.36]{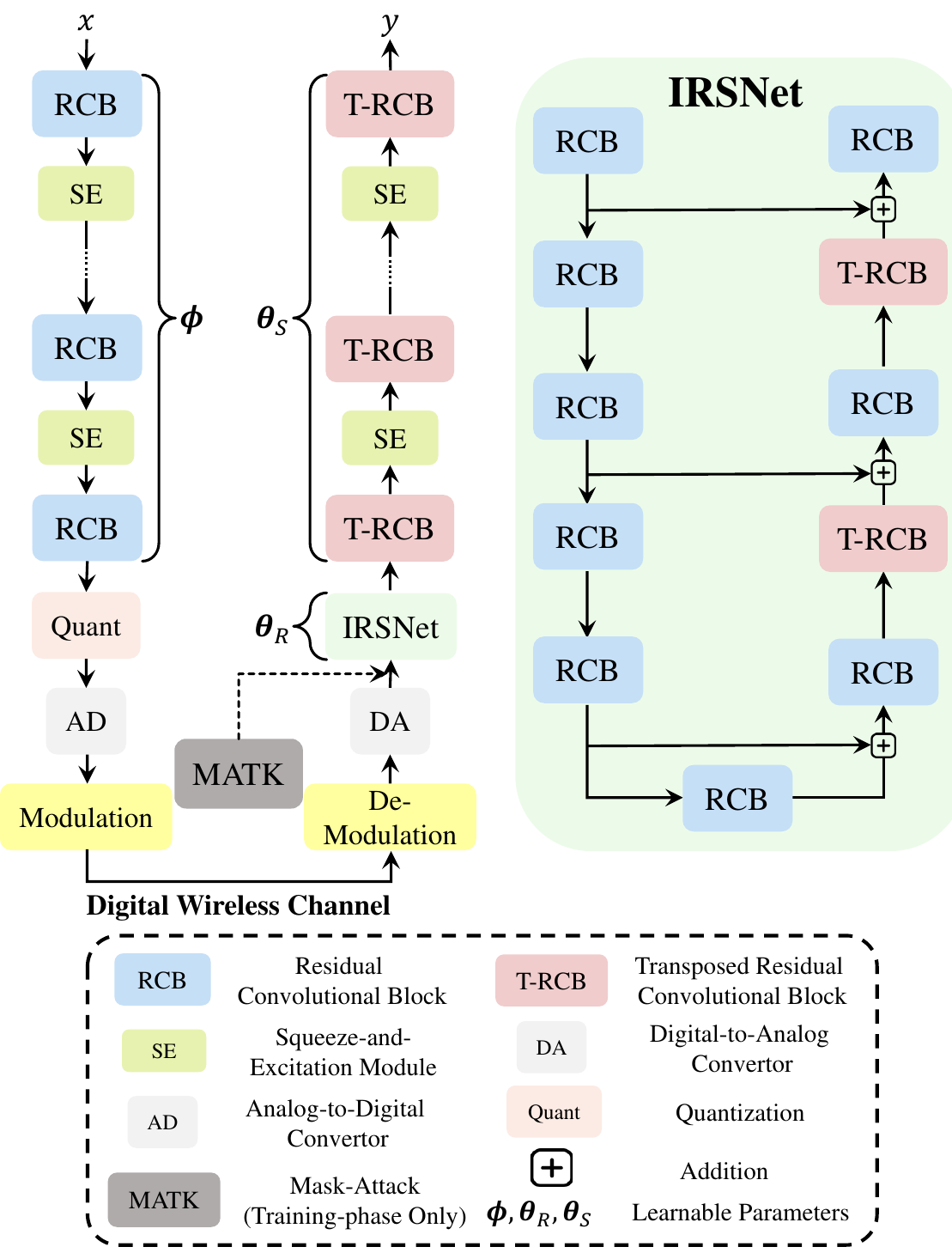}
    \caption{\small{The proposed digital SemComm system for image transmission. Left: The pipeline of AMP-SC; Right: The information restoration network IRSNet; Below: Notations. }}
    \label{fig:proposed_system}
  \end{figure}

  The network architecture of proposed digital SemComm system is shown in Fig.~\ref{fig:proposed_system}. 
  The learnable part of the system includes multiple residual convolutional blocks (RCBs) \cite{he2016deep}, transposed residual convolutional blocks (T-RCBs), and squeeze-and-excitation modules (SEs)\cite{hu2018squeeze}. 
  Notably, we employ two convolutional layers in RCB and two transposed convolutional layers in T-RCB respectively. 
  Assuming the input and output of modules are $\bm{s}$ and $\bm{s}^{'}$, the processing function of RCB and T-RCB can be given by 
  \begin{equation}
    \bm{s}^{'} = \mathcal{F}_1(\bm{s}, \{W_i\}) + W_s\bm{s},
  \end{equation}
  where $\mathcal{F}_1(\bm{s}, \{W_i\})$ represents the learnable residual mapping function with parameter $\{W_i\}$ and consists of two convolution operations. 
  In addition, $W_s$ is denoted as the parameters of the shortcut convolutional layer. 
  The processing function of SE is given by 
  \begin{equation}
    \bm{s}^{'} = \mathcal{F}_2(\bm{s}, \{W_i\}) \times \bm{s},
  \end{equation}
  where $\mathcal{F}_2(\bm{s}, \{W_i\})$ represents squeeze-and-excitation function and contains two linear operations and a sigmoid function. 
  The ``Squeeze'' operation in the SE compresses features in each channel through global average pooling, while the ``Excitation'' operation enhances the representation of informative features using learned channel attention weights. 
 
  The proposed IRSNet has an architecture similar with U-Net, which is a widely used framework in image restoration. 
  As shown in the middle of Fig.~\ref{fig:proposed_system}, IRSNet consists of nine RCBs and two T-RCBs. 
  Similar with the general U-Net architecture, the first five RCBs are utilized to down-sample the input features, and one RCB for projection.
  In the up-sampling stage, three RCBs and two T-RCBs are cross-used to recover the transmitted semantic information. 
  Since ``fake'' semantic information introduced by BFP represents nothing in the source data, it will be detected and restored in higher-dimensional latent space through down-sampling and up-sampling processes, respectively.  
  Accordingly, the distortion caused by BFP remains minimal in the restored signals. 

  \begin{table}[!t]
    \caption{\small{Training Settings of AMP for AMP-SC}}
    \centering
    \resizebox{0.9\linewidth}{!}{
      \begin{tabular}{|c|c|c|c|}
      \hline
      Steps & Epoch  & Learning Rate & Mask Ratio\\
      \hline
      $1$     & $4T$   & $1e-4$ & N/A\\
      \hline
      $2$     & $T$   & $1e-5$ & N/A\\
      \hline
      $3$     & [$4T$, $T$, $T$] & [$1e-4$, $1e-4$, $1.5e-5$] & [$2p$, $2p$, $2p$]\\
      \hline
      $4$     & [$T$, $T$, $T$] & [$1e-4$, $1e-4$, $1.5e-5$] & [$2p$, $2p$, $2p$]\\
      \hline
      $5$     & $T$   & $1.5e-5$ & N/A\\
      \hline
      \end{tabular}%
    \label{tab:training_settings_ours}%
    }
  \end{table}%
\section{Simulation Results}\label{simulation_results}
    In this section, we evaluate the image reconstruction performance of the proposed AMP-SC and benchmarks.
\subsection{Experiment Settings}
    \subsubsection{Dataset} 
    The simulations are performed on three datasets in different scales, including CIFAR10, Kodak24, and DIV2K. 
    CIFAR10 dataset consists 50,000 R.G.B images at the size of $3 \times 32 \times 32$ for training and 10,000 images for test. 
    To further evaluate the superiority of AMP-SC, we introduce Kodak24 dataset, which includes 24 R.G.B. images at the size of $3 \times 512 \times 768$ or $3 \times 768 \times 512$, and DIV2K dataset, whose validation part consists 80 R.G.B. images with approximate 2K resolution, for validation on high-resolution images.



\subsubsection{Benchmarks}
    To demonstrate the superiority of the proposed AMP-SC and AMP, we introduce the following benchmarks, which are highly related to our work, for comparsion:
    \begin{itemize}
        \item \textbf{Conventional}: This benchmark is a performance-achieving conventional SemComm system, whose parameters are from the converged model in Step-1 of the AMP.  Moreover, it is evaluated under zero BFP ratio condition.
        \item \textbf{SoftQ}: Similar to the approach in \cite{shao2023task}, we adopt a digital SemComm system which is trained with additive noise in (\ref{y2}) to achieve ``soft'' quantization in the analog domain. In other words, the SoftQ model represents the converged model in Step-2 of the AMP.
        \item \textbf{He22}: We construct a digital system with an architecture identical to that of AMP-SC without IRSNet. Following the two-phase training strategy in \cite{he2022robust}, we train the system with FE and TTA in Fig.~\ref{fig:pipeline} to serve as a benchmark. 
        \item \textbf{SoftHe22}: We fine-tune the decoder networks in SoftQ with the second training phase in He22. 
    \end{itemize}
    For fair comparison, the ratio of the number of transmitted continuous symbols and that of pixels in the source image is fixed at $1/6$. 

    \begin{table}[!t]
        \centering
        \caption{\small{Number of Blocks}}
        \begin{tabular}{|c|c|c|c|c|}
        \hline
        \multirow{2}[0]{*}{Block} & \multicolumn{2}{c|}{CIFAR10} & \multicolumn{2}{c|}{Kodak24 \& DIV2K} \\
    \cline{2-5}          & Encoder & Decoder & Encoder & Decoder \\
        \hline
        RCB / T-RCB & 6     & 6     & 8     & 8 \\
        \hline
        SE    & 5     & 5     & 7     & 7 \\
        \hline
        \end{tabular}%
        \label{tab:number_of_blocks}%
    \end{table}%

    \begin{table}[!t]
        \centering
        \caption{\small{BFP Ratio versus SNR}}
        \resizebox{0.95\linewidth}{!}{
          \begin{tabular}{|c|c|c|c|c|c|c|c|}
          \hline
          SNR(dB) & 0     & 1     & 2     & 3     & 4     & 5     & 6 \\
          \hline
          $p$     & 1.41e-01 & 1.19e-01 & 9.77e-02 & 7.75e-02 & 5.86e-02 & 4.19e-02 & 2.79e-02 \\
          \hline
          SNR(dB) & 7     & 8     & 9     & 10    & 11    & 12    & 13 \\
          \hline
          $p$     & 1.70e-02 & 9.25e-03 & 4.39e-03 & 1.75e-03 & 5.65e-04 & 1.39e-04 & 2.42e-05 \\
          \hline
          SNR(dB) & 14    & 15    & 16    & 17    & 18    &       &  \\
          \hline
          $p$     & 2.76e-06 & 1.84e-07 & 6.25e-09 & 9.07e-11 & 4.52e-13 &       &  \\
          \hline
          \end{tabular}%
        \label{tab:bfr}%
        }
    \end{table}%

\subsubsection{Implementation Details}
    The number of blocks in the encoder and decoder for all evaluated systems is shown in Table~\ref{tab:number_of_blocks}. 
    The training configuration for the proposed AMP-SC and all benchmarks is outlined as follows.
    We use the Adam stochastic optimization method to update the model parameters.
    Additionally, a cosine annealing strategy is introduced to dynamically adapt the learning rate to achieve better training. 
    For modulation and demodulation, we respectively utilize 16-QAM and Gray code.
    Without loss of generality, following \cite{guo2023device,bao2024sdac,he2022robust, choi2019neural, park2023joint, song2020infomax}, we implement the modulation-demodulation process over an AWGN channel by going through the bit-stream with BFP effect.
    Notably, the projection from SNR to the flipping probability $p$ in AWGN channel is shown in Table~\ref{tab:bfr}, which is generated by the Bit-Error-Rate-Analysis toolbox provided by MATLAB.
    To train AMP-SC, the $p$ in AMP is set as $0.0125$. 
    Moreover, the alternating training lasts for three rounds\footnote{
        Although increasing the number of alternating training rounds can gradually improve task performance, it is out of scope of this paper. 
        Additionally, considering the efficiency of simulations, we set the number of rounds to three without loss of generality.
    }. 
    In addition, the intensity coefficient $\alpha$ of additive randomness $\bm{\mathcal{N}}$ is set as 0.5.

    For simulations on CIFAR10 dataset, all systems are trained and tested on the dataset, and the batch size is set as 256. 
    In particular, the training hyperparameter $T$ in Table.~\ref{tab:training_settings_ours} is set as 200. 
    On the other hand, for simulations on high-resolution images, we follow \cite{zhang2023predictive} to leverage the validation set of ImageNet dataset, including 48627 R.G.B. images, to train the systems and test them on Kodak24 and DIV2K datasets. 
    For each training sample, it is obtained by random-crop the original images with size $256 \times 256$, the batch size is set as 16, and the training hyperparameter $T$ is set as 15. 

    \begin{figure}[!t]
        \centering
        \begin{subfigure}{0.85\linewidth}
            \centering
            \includegraphics[width=\linewidth]{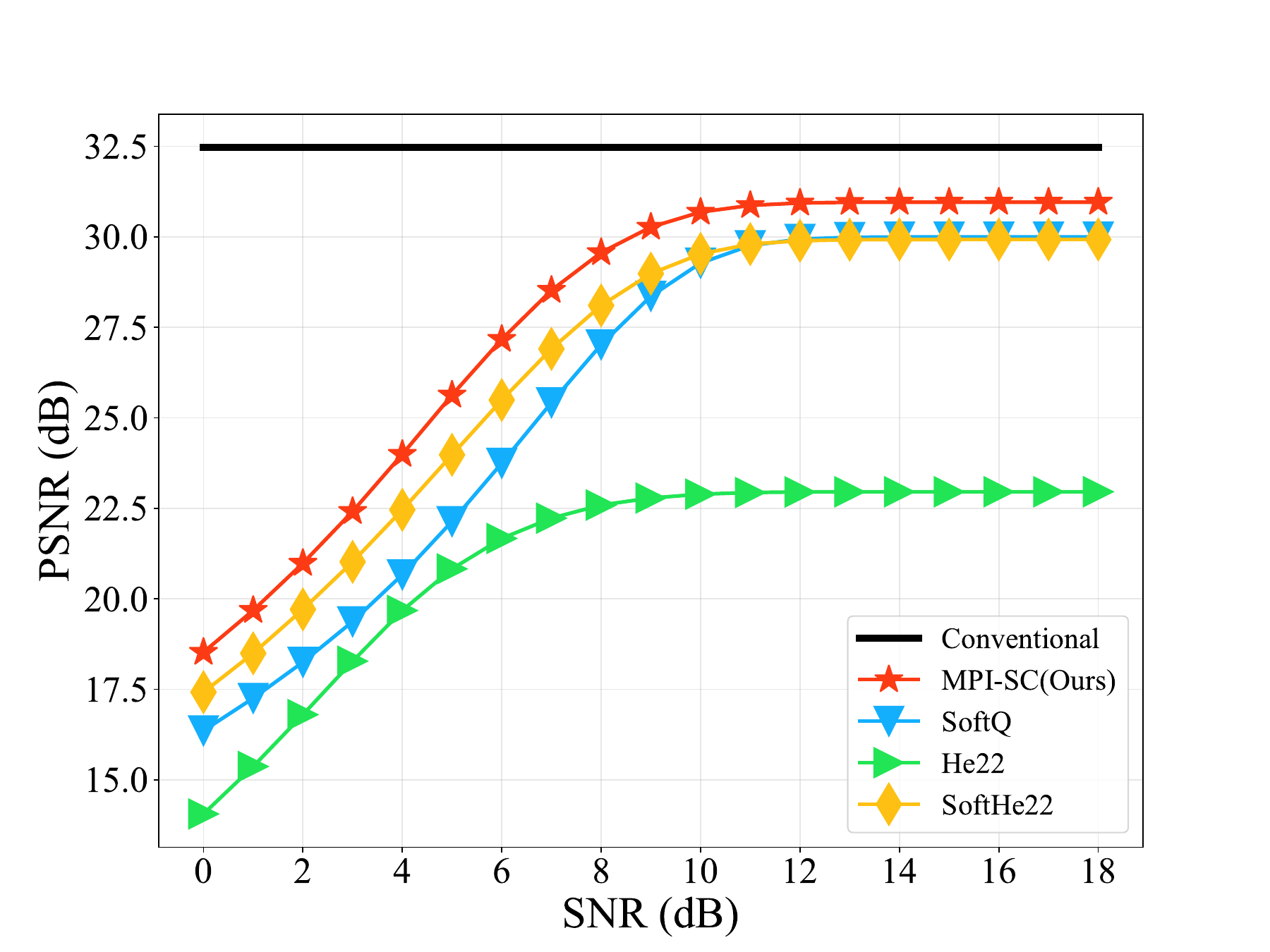}
            \caption{4-bits}
            \label{fig:cifar10_4bits}
        \end{subfigure}
        \begin{subfigure}{0.845\linewidth}
            \centering
            \includegraphics[width=\linewidth]{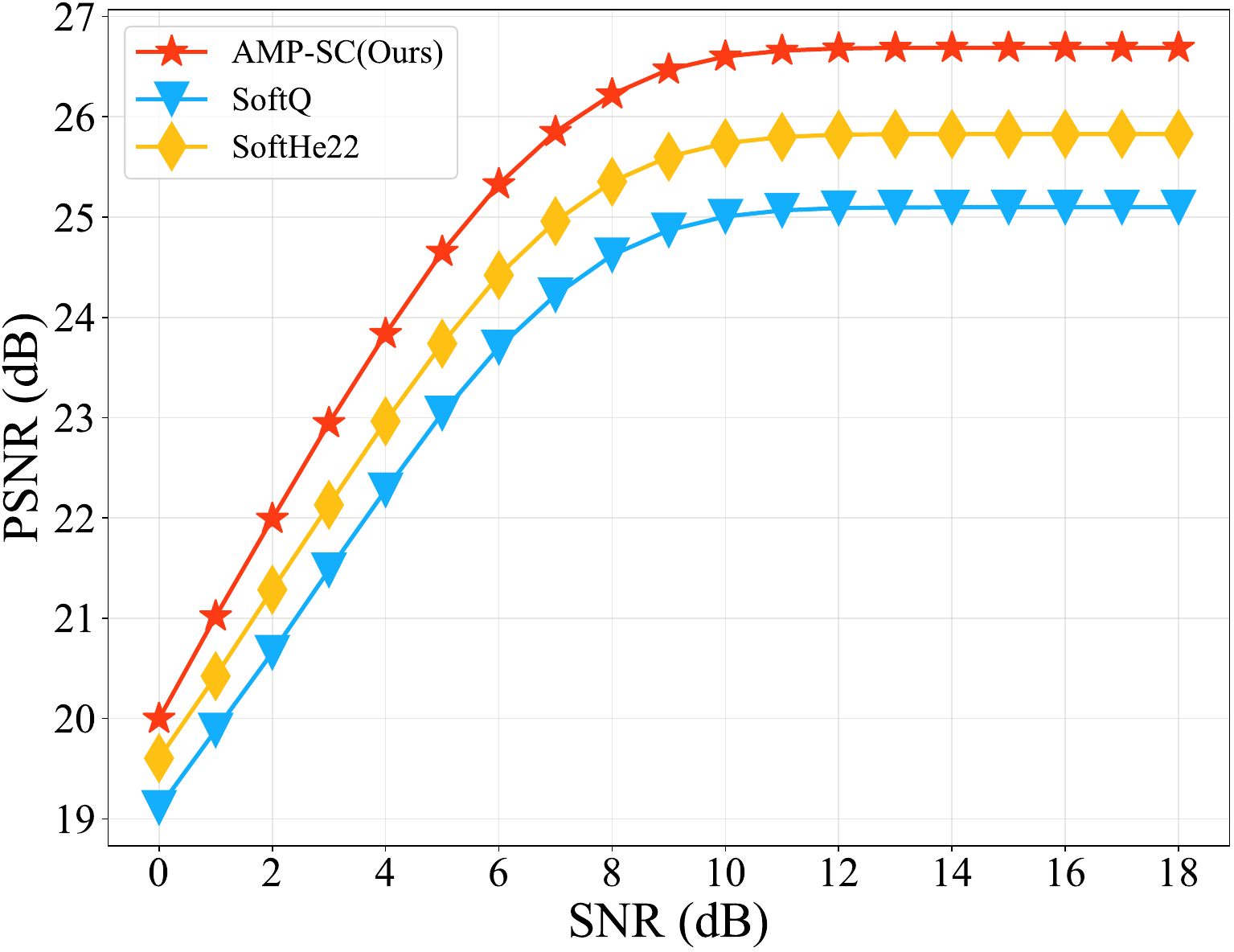}
            \caption{2-bits}
            \label{fig:cifar10_2bits}
        \end{subfigure}
        \caption{\small{The PSNR of AMP-SC and benchmarks in an AWGN channel versus SNR from 0dB to 18dB in CIFAR10 dataset. Notably, each transmitted symbol is converted into 4 or 2 bits.}}
        \label{fig:cifar10}
    \end{figure} 

\subsection{Validation Results}
\subsubsection{Results on CIFAR10 Dataset}
    In this experiment, we evaluate the reconstruction quality achieved by the proposed AMP-SC and benchmarks in an additive white Gaussian noise (AWGN) channel versus a wide range of SNR in CIFAR10 dataset. 
    In Fig.~\ref{fig:cifar10_4bits}, we plot the achieved peak signal-to-noise ratio (PSNR) of total five systems, where are deployed 4-bit analog-digital conversion.
    Overall, AMP-SC displays less performance degradation across all SNRs compared to the baseline and benchmarks, showcasing its superiority among digital SemComm systems.
    On average, AMP-SC achieves PSNR of $27.63$dB in reconstructing source images, which is $1.97$dB higher than that of SoftQ. 
    In particular, AMP-SC demonstrates $6.53$dB and $1.24$dB higher PSNR than those achieved by He22 and SoftHe22, respectively.
    These improvements highlight the effectiveness of the proposed AMP and IRSNet. 
    Although the reproduced two-phase training strategy is not so superior as that reported in \cite{he2022robust}, we attribute this difference to potential implementation gaps due to variations in network architectures.
    Comparing SoftQ and SoftHe22, the higher performance of SoftHe22 illustrates the effectiveness of the two-phase training strategy in \cite{he2022robust}. 
    \emph{For fair comparison, we utilize SoftHe22 to represent the benchmark based on two-phase training strategy, rather than He22, in the following. }

    \begin{figure}[!t]
        \centering
        \begin{subfigure}{0.852\linewidth}
            \centering
            \includegraphics[width=\linewidth]{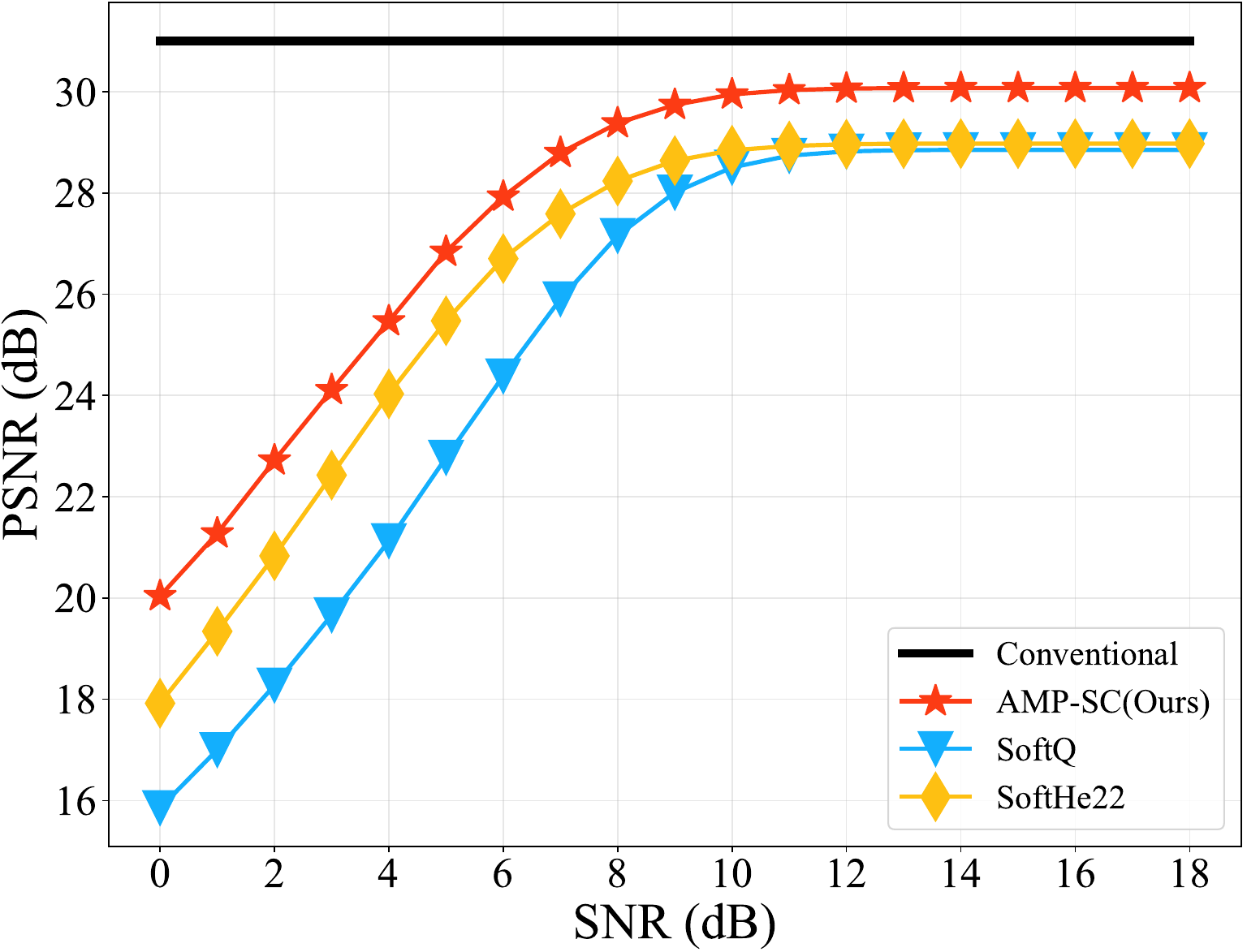}
            \caption{4-bits}
            \label{fig:kodak24_4bits}
        \end{subfigure}
        \begin{subfigure}{0.872\linewidth}
            \centering
            \includegraphics[width=\linewidth]{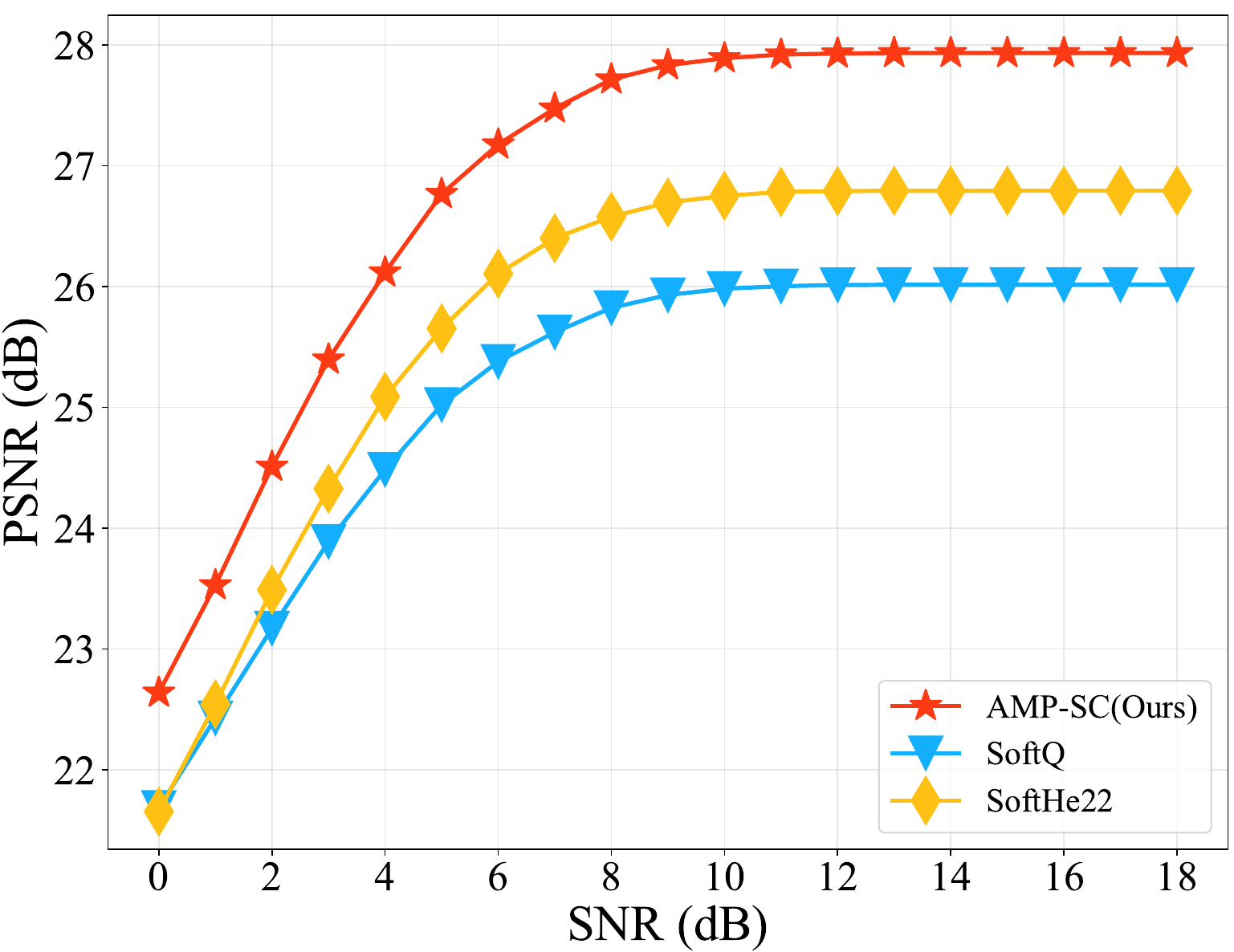}
            \caption{2-bits}
            \label{fig:kodak24_2bits}
        \end{subfigure}
        \caption{\small{The PSNR of AMP-SC and benchmarks in an AWGN channel versus SNR from 0dB to 18dB in Kodak24 dataset. Notably, each transmitted symbol is converted into 4 or 2 bits.}}
        \label{fig:kodak24}
    \end{figure}

    \begin{figure}[!t]
        \centering
        \begin{subfigure}{0.88\linewidth}
            \centering
            \includegraphics[width=\linewidth]{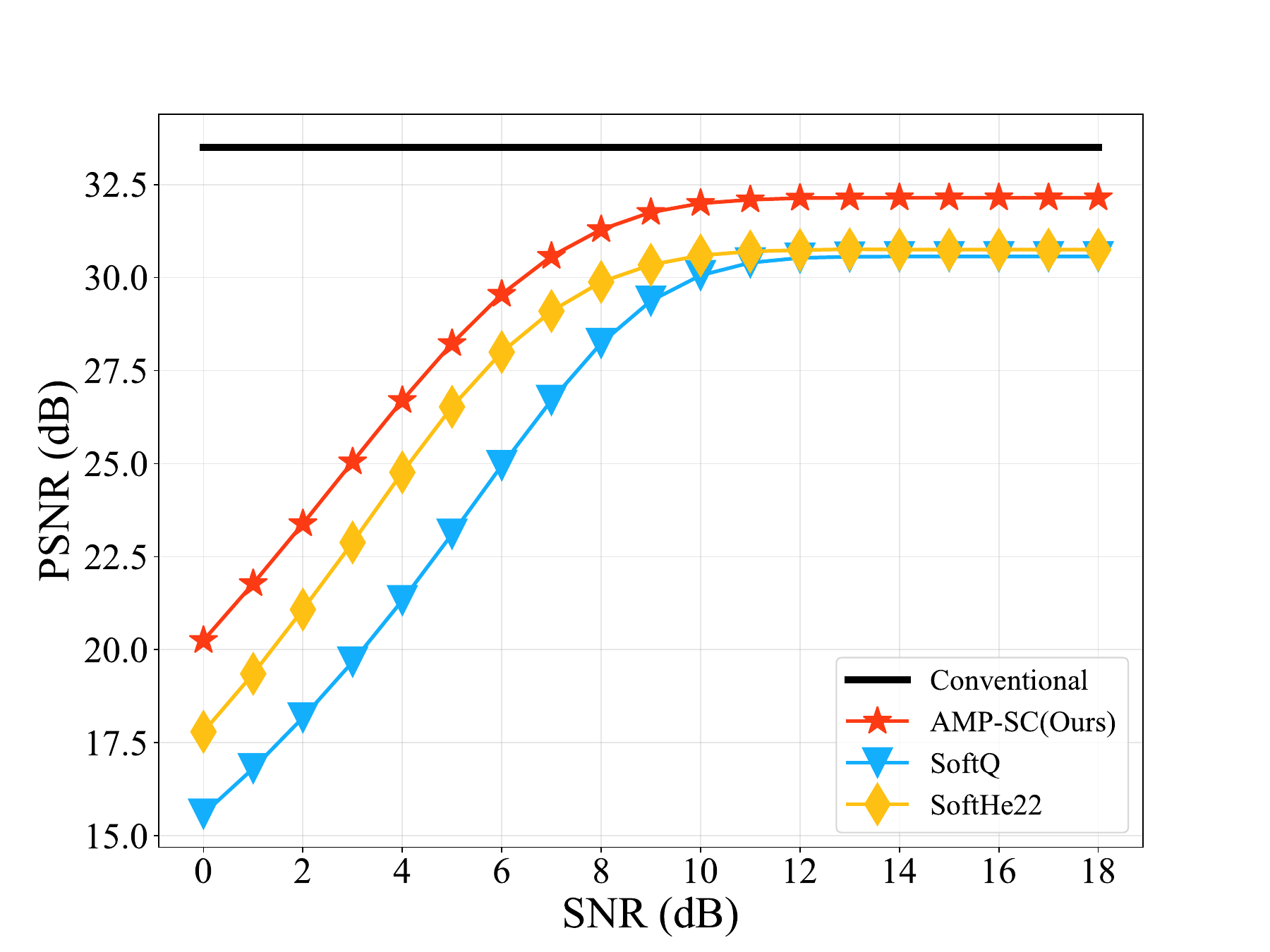}
            \caption{4-bits}
            \label{fig:div2k_4bits}
        \end{subfigure}
        \begin{subfigure}{0.872\linewidth}
            \centering
            \includegraphics[width=\linewidth]{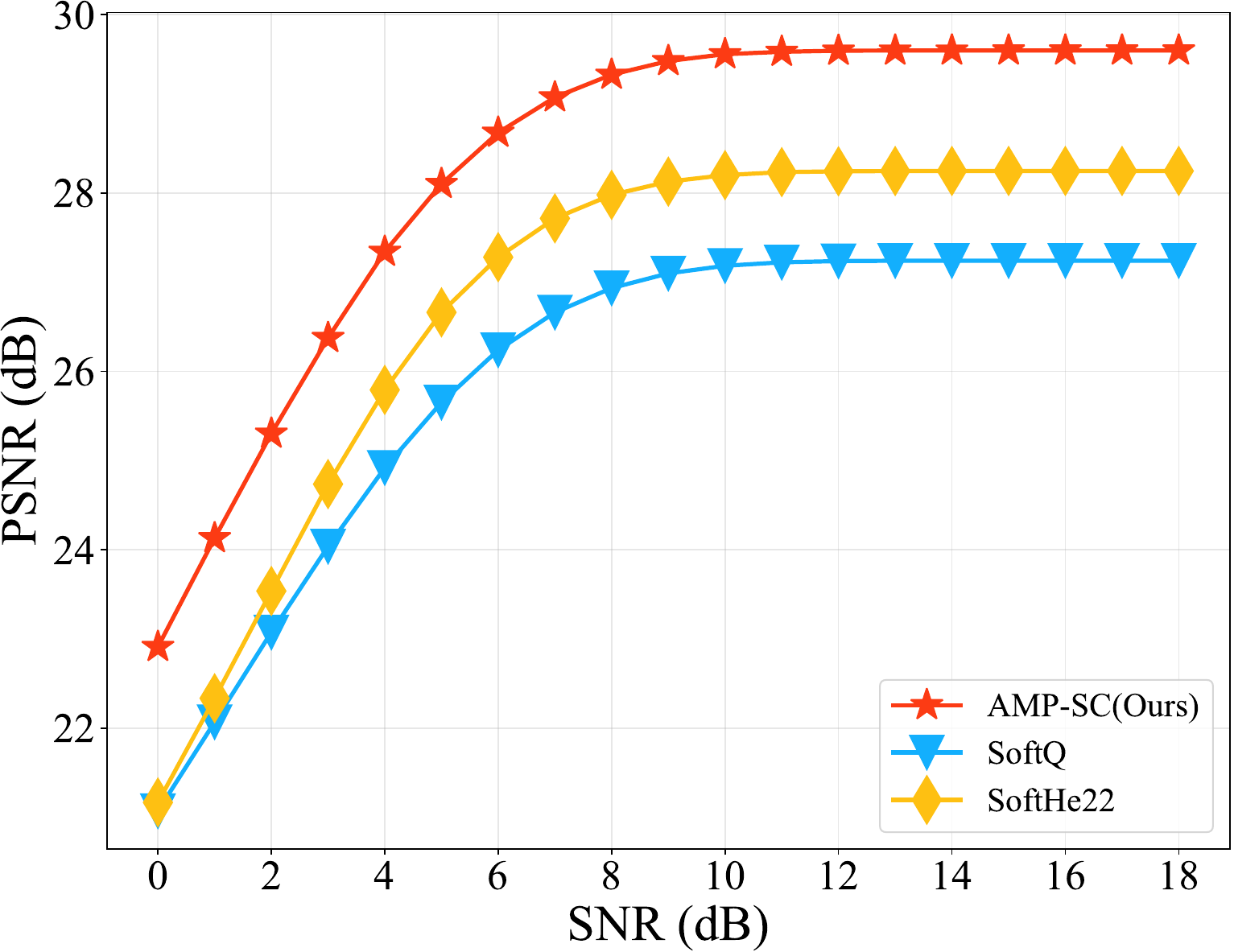}
            \caption{2-bits}
            \label{fig:div2k_2bits}
        \end{subfigure}
        \caption{\small{The PSNR of AMP-SC and benchmarks in an AWGN channel versus SNR from 0dB to 18dB in DIV2K dataset. Notably, each transmitted symbol is converted into 4 or 2 bits.}}
        \label{fig:div2k}
    \end{figure} 

    \begin{figure*}[!t]
        \centering
        \includegraphics[scale=0.66]{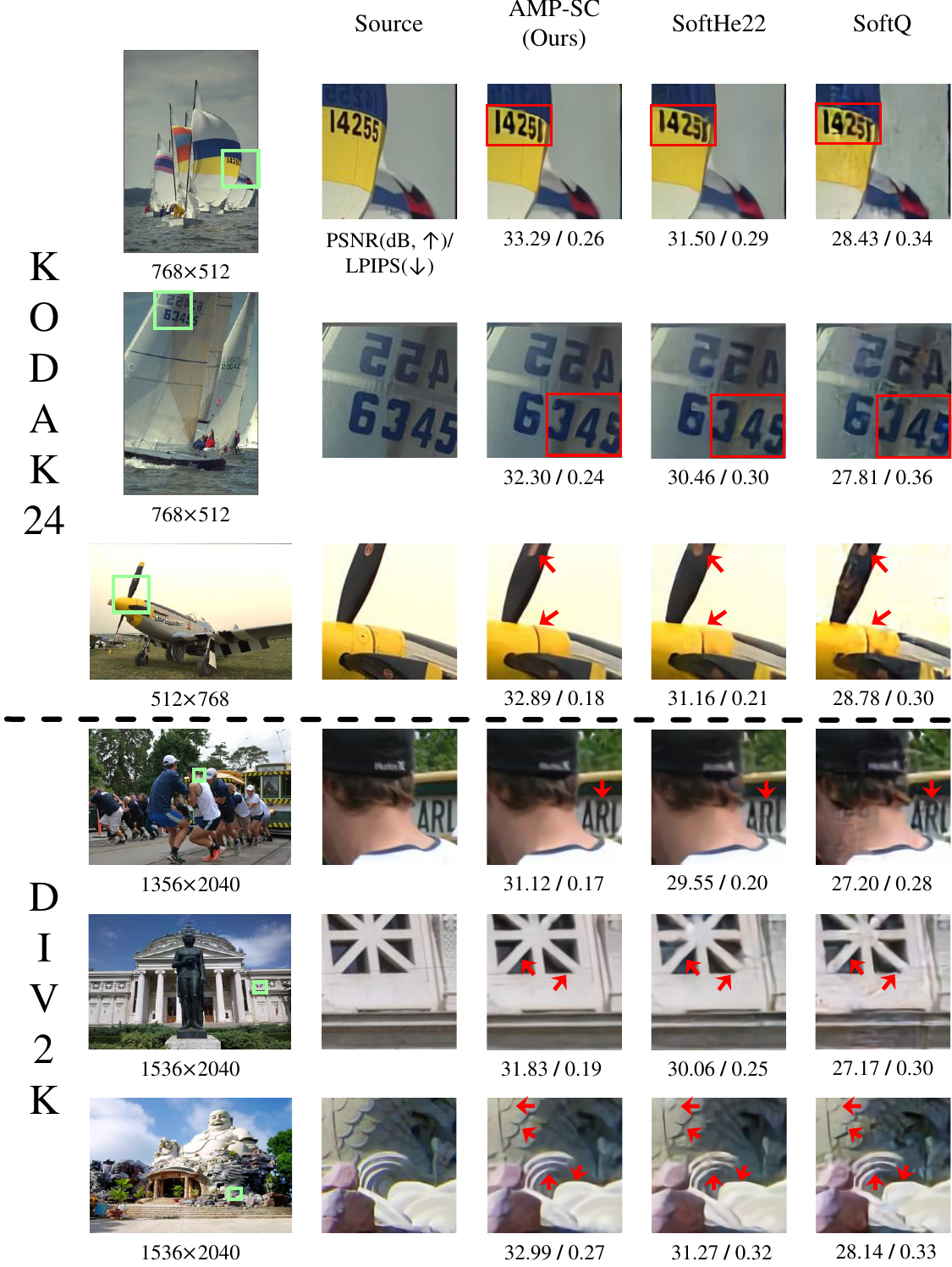}
        \caption{\small{Visualization Results of AMP-SC and two representative benchmarks in Kodak24 and DIV2K datasets under AWGN channel at SNR=7dB. 
        For better representation, we zoom part of the image with a $128 \times 128$ window to demonstrate the differences in details. }}
        \label{fig:visualization_results}
    \end{figure*} 

    Additionally, in Fig.~\ref{fig:cifar10_2bits}, we evaluated AMP-SC, SoftQ, and SoftHe22, where each transmitted symbol is converted into 2 bits. 
    Notably, compared to SoftHe22, AMP-SC achieves an average performance improvement of $0.82$dB, highlighting the better noise robustness enhanced by the proposed AMP and IRSNet.
    Overall, the simulation results in Fig.~\ref{fig:cifar10} confirm the superiority of the proposed AMP-SC and highlight the robustness, enhanced by the combination of AMP and IRSNet, against information loss in digital operations.

\subsubsection{PSNR Performance on High-Resolution Datasets}
    In the following, we leverage two datasets with high-resolution images to further evaluate the reconstruction performance of the proposed system and benchmarks. 
    Generally, as shown in Fig.~\ref{fig:kodak24} and Fig.~\ref{fig:div2k}, the proposed AMP-SC still holds the lead in digital image transmission. 
    Notably, the advantage in noise robustness of AMP-SC becomes more significant when the channel condition is poor. 
    For example, under the setting of 2-bit analog-digital conversion, AMP-SC achieves $0.39$dB higher PSNR across AWGN channel with intensity of 0dB in CIFAR10 dataset. 
    However, in Kodak24 and DIV2K datasets, AMP-SC displays $0.98$dB and $1.74$dB higher PSNR than SoftHe22, respectively. 
    When the number of bits for conversion increases to 4, AMP-SC achieves $2.12$dB and $2.46$dB higher reconstruction performance than SoftHe22 in Kodak24 and DIV2K datasets, respectively, while it illustrates $1.10$dB less performance degradation in CIFAR10 dataset. 

    The above observation supports the advantage of AMP-SC in reconstructing detailed source information.
    It is noteworthy that, from CIFAR10 to Kodak24 and DIV2K datasets, the detailed information of each unit block in the source image, such as texture, becomes richer. 
    Richer details place higher requirements on transmission fidelity. 
    Given poor channel environment, compared with the relative performance improvement of AMP-SC in CIFAR10 dataset, those in high-resolution datasets demonstrate the superiority of AMP-SC in high-fidelity image transmission in digital fashion.
    In addition, AMP-SC also enjoys the remarkable compatibility with current wireless communication framework. 

\subsubsection{Visualization Results}
    To further demonstrate the effectiveness of our proposed system, we provide a set of visually intuitive results on high-resolution datasets in Fig.~\ref{fig:visualization_results}. 
    The visual comparisons are conducted under AWGN channel with noise intensity of 7dB. 
    In addition, we introduce the learned perceptual image patch similarity (LPIPS) metric to provide an additional perspective for evaluating reconstruction quality.
    For better representation, part of the image is zoomed in by a window, whose size is $128 \times 128$, to show the details. 
    Therefore, the differences in details can be revealed, showing the performance gap between the systems more intuitively. 
    In brief, from these visualization results, we can observe that our proposed AMP-SC exhibits better visual quality over the digital channel. 

    As we mentioned before, it is important for the high-resolution images to keep detailed information during transmission as much as possible for visual experience. 
    On the other words, the superiority of digital SemComm systems can be determined by the details in reconstructed images. 
    As shown in the highlighted parts of Fig.~\ref{fig:visualization_results}, SoftQ reconstructs high-resolution images with blur details and slight watermark effect, leaving negative impact to visual experience. 
    With network fine-tuning in the decoder, SoftHe22 effectively improves sharpness of the reconstructed images. 
    However, SoftHe22 demonstrates weakness in recovering detailed information.
    For example, in the third row of Fig.~\ref{fig:visualization_results}, the seam between the two parts of the aircraft that reconstructed by SoftHe22 is incomplete. 
    Moreover, in the forth row, the letter ``R'' in the reconstructed image has a similar issue. 
    Beyond SoftHe22 and SoftQ, AMP-SC displays images with more authentic detailed information and higher sharpness in the receiver side, offering better visual experience to the user. 
    Overall, both the comparisons and visualization results in high-resolution datasets support the superiority of the proposed AMP-SC.

\subsubsection{Ablation Study on Proposed Component}
    In this ablation study, we investigate the effectiveness of the proposed module and training strategy in DIV2K dataset. 
    With respect to ``soft'' quantization in \cite{shao2023task} and the two-phase training strategy proposed by \cite{he2022robust}, we construct a digital SemComm system based on ``soft'' quantization and the proposed AMP, which is named SoftAMP. 
    Notably, since IRSNet are not introduced in SoftAMP, for all steps in AMP, the training epochs are set as $T$. 
    To further validate the effectiveness of the introduced MATK, we consider two more benchmarks, named SoftHe22-wMATK and SoftAMP-woMATK.
    Specifically, we introduce MATK to SoftHe22 in the second training phase to construct SoftHe22-wMATK.
    Additionally, we remove MATK from SoftAMP during training to construct SoftAMP-woMATK.
    
    \begin{figure}[!t]
        \centering
        \includegraphics[scale=0.27]{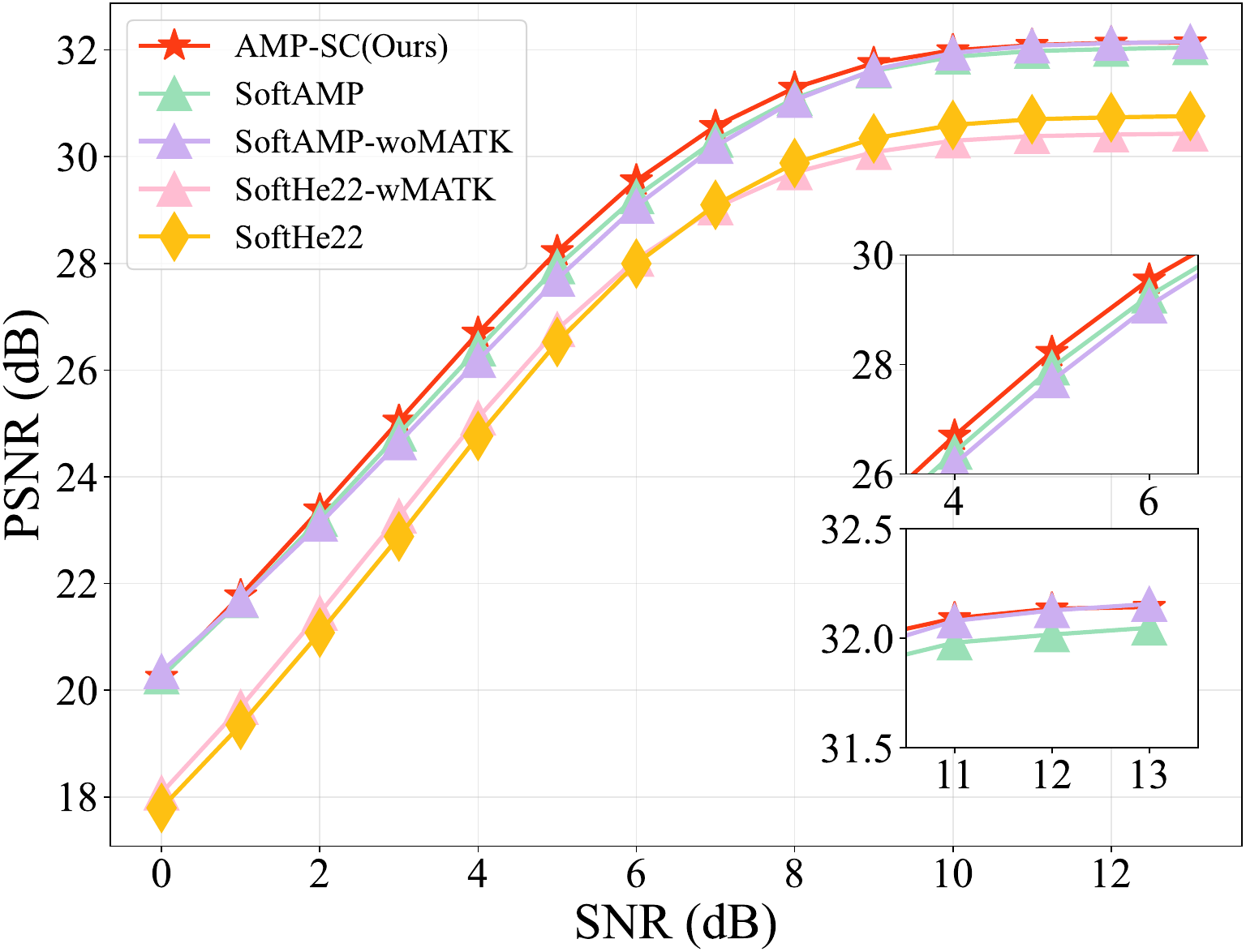}
        \caption{\small{Inference performance versus SNR from 0dB to 13dB on DIV2K dataset. Each transmitted symbol is converted into 4 bits. Notably, SoftAMP is equivalent to removing IRSNet from AMP-SC in network architecture. }}
        \label{fig:ablation_study}
    \end{figure} 

    As shown in Fig.~\ref{fig:ablation_study}, all of AMP-SC, SoftAMP, and SoftAMP-woMATK display higher inference performance than SoftHe22 under the 4-bit analog-digital conversion over AWGN channel with a wide range of noise intensity. 
    In other words, the proposed AMP supports digital SemComm system in image transmission better than the two-phase training strategy, evaluating the superiority of AMP. 
    Furthermore, although AMP-SC and SoftAMP hold the close lower and upper bounds of reconstruction performance over AWGN channel, AMP-SC demonstrate higher inference accuracy than SoftAMP from 1dB to 10dB. 
    It gives the credit to the proposed IRSNet. 
    In addition, given an image with size of $3 \times 128 \times 128$ as the input, the number of parameters (Param) and floating point operations (FLOPs) of in the receiver are shown in Table~\ref{tab:flops_param}.  
    It is worth noting that the extra storage and computation overhead introduced by IRSNet are not significant, while IRSNet further improves the noise robustness in the receiver.


    
    Notably, by introducing MATK, SoftHe22-wMATK demonstrates higher performance than SoftHe22 from 0dB to 7dB, while the upper bound of SoftHe22-wMATK is less than that of SoftHe22.
    Additionally, the curves of SoftAMP and SoftAMP-woMATK hold a similar relationship, but the gap between their upper bounds is smaller than that between SoftHe22-wMATK and SoftHe22.
    Above observation supports the effectiveness of MATK in improving robustness to digital noise and the necessity of TTA in AMP.
    Overall, the effectiveness of the proposed IRSNet and MATK is validated. 

    \begin{figure}[!t]
        \centering
        \includegraphics[scale=0.27]{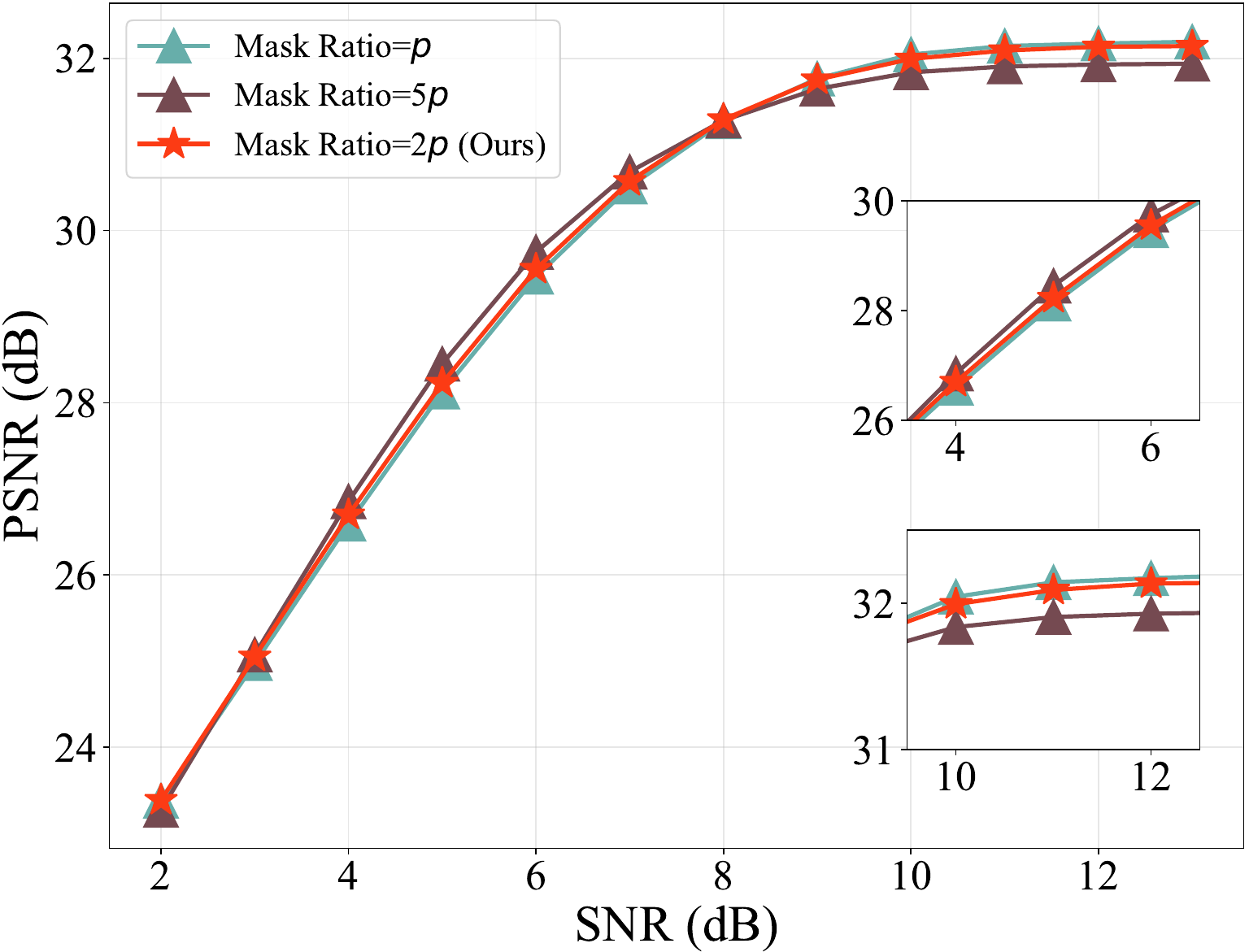}
        \caption{\small{Inference performance versus SNR from 2dB to 13dB on DIV2K dataset. Notably, each transmitted symbol is converted into 4 bits.}}
        \label{fig:mask_ratio}
    \end{figure} 

    \begin{table}[!t]
        \centering
        \caption{\small{Number of parameters (Param) and floating point operations (FLOPs) number of parameters in the receiver}}
        \begin{tabular}{|c|c|c|}
        \hline
        System    & Ours & SoftHe22 \\
        \hline
        Param (M) & 12.917 & 11.180 \\
        \hline
        FLOPs (G) & 30.834 & 30.805 \\        
        \hline
        \end{tabular}%
        \label{tab:flops_param}%
    \end{table}%

\subsubsection{Ablation Study on Mask Ratio}
    In this ablation study, we investigate the mask ratio from $p$ to $5p$ in DIV2K dataset, where $p$ is set to $0.0125$ during training. 
    As shown in Fig.~\ref{fig:mask_ratio}, setting mask ratio to $2p$ can achieve a balance between noise robustness and representation ability of the digital SemComm system. 
    Notably, the relative relationship changes between mask ratios $p$ and $5p$ as the SNR increases. 
    This observation highlights the defects of these two training settings of mask ratio in terms of robustness and representation ability. 
    
\section{Conclusions}\label{conclusion}
    In this paper, we propose an alternating multi-phase training strategy (AMP) to overcome the non-differentiability inherent in bits transmission.     
    In the first phase, the system is trained in the analog domain to extract the semantic-relevant information and compress the irrelevant information.
    Towards the compatibility to current digital communication framework, we enhance system robustness against information loss caused by digital operations in the second phase.
    In the third phase, we bridge the gap between training and testing to further improve task performance. 
    Moreover, we investigate a MATK in the second phase to simulate an evident and severe BFP effect in a differentiable manner. 
    Combining with MATK, we also developed a IRSNet to improve inference accuracy by restoring the demodulated signals. 
    Additionally, we proposed a digital semantic communication system named AMP-SC for image transmission, which is compatible with current digital communication framework. 
    Simulation results have illustrated the outstanding reconstruction performance of the proposed AMP-SC. 
    Particularly, in the context of 4-bit quantization, AMP-SC achieves $1.24 \sim 1.65$dB higher PSNR than the system, which is empowered by the ``soft'' quantization and two-phase training strategy, on average among three representative datasets and a wide range of SNR.

\bibliography{ref}    
\bibliographystyle{ieeetr}
\end{document}